\documentclass[%
 prd,
 jmp,%
 reprint,%
 onecolumn,
 superscriptaddress,
,12pt
]{revtex4-2}

\usepackage{amssymb}
\usepackage{amsmath}
\usepackage{graphicx}
\usepackage{dcolumn}
\usepackage{bm}
\usepackage{color}
\usepackage{xcolor}
\usepackage{float}
\allowdisplaybreaks
\usepackage{subfig}
\usepackage{comment}
\usepackage[figurename=FIGURE,tablename=TABLE]{caption}
\usepackage{threeparttable}
\usepackage{hyperref}
\hypersetup{
    colorlinks=true, 
    linktoc=all,     
    linkcolor=magenta,
    citecolor=blue
}

\begin{document}

\title{Dark Matter and Dark Energy from a Kaluza-Klein inspired Brans-Dicke Gravity with Barotropic Fluid}

\author{Areef Waeming}
\email{reef.waeming@gmail.com}
\affiliation{Khon Kaen Particle Physics and Cosmology Theory Group (KKPaCT), Department of Physics, Faculty of Science, Khon Kaen University, 123 Mitraphap Rd., Khon Kaen, 40002, Thailand}
\author{Tanech Klangburam}%
\email{klangburam.t@gmail.com}
\author{\\Chakrit Pongkitivanichkul}
\email{chakpo@kku.ac.th}
\author{Daris Samart}
\email{darisa@kku.ac.th}
\affiliation{Khon Kaen Particle Physics and Cosmology Theory Group (KKPaCT), Department of Physics, Faculty of Science, Khon Kaen University, 123 Mitraphap Rd., Khon Kaen, 40002, Thailand}
\affiliation{National Astronomical Research Institute of Thailand, Chiang Mai 50180, Thailand}


\date{\today}

\begin{abstract}

We study the Kaluza-Klein inspired Brans-Dicke model with barotropic matter. Following from our previous work, the traditional Kaluza-Klein gravity action is introduced with an additional scalar field and 2 gauge fields. The compactification process results in a Brans-Dicke model with a dilaton coupled to the tower of scalar fields whereas a gauge field from 5-dimensional metric forms a set of  mutually orthogonal vectors with 2 additional gauge fields. The barotropic matter is then introduced to complete a realistic set up. To demonstrate the analytical solutions of the model, we consider the case in which only 2 lowest modes becoming relevant for physics at low scale. After derivation, equations of motion and Einstein field equations form a set of autonomous system. The dynamical system is analysed to obtain various critical points. Interestingly, by only inclusion of barotropic matter, the model provides us the critical points which capable of determining the presences of dark matter, dark energy and phantom dark energy.
\end{abstract}

\maketitle

\newpage
\section{\label{sec:Introduction}Introduction}
The origin of dark matter (DM) and dark energy (DE) has been one of the main pursuits for new physics for many decades. DM was proposed as an invisible mass that holds a galaxy together while DE was an explanation of the accelerating universe. Although there are many candidates from the particle physics side such as Axion-Like Particle (ALP) which could act as both DM and DE depending on their masses \cite{Arvanitaki:2009fg,Marsh:2015xka}, but none of these have been experimentally confirmed yet.
On the other hand, a natural place for which the candidate for DM and DE could rise is the gravity theory. Over the past decades, a tremendous progress has been made in generalisation of Einstein gravity especially in the class of models called the Scalar-Tensor theory \cite{Fujii:2003pa, Faraoni:2004pi} and reference therein. 
One of the most studied and celebrated models is Brans-Dicke (BD) theory \cite{Brans:1961sx}. Motivated by Mach's principle, the scalar field is included to modify the gravity in a way that Weak Equivalence Principle is still intact. As a result, the gravitational constant is promoted to a scalar field which coupled to both mass and geometry. A huge number of studies for this particular model can be found here \cite{Kolitch:1994kr,Santos:1996jc,Copeland:1997et,Abdalla:2007qi,deSouza:2005ig,Hrycyna:2013hla,Hrycyna:2013yia,Hrycyna:2014cka,Garcia-Salcedo:2015naa,Papagiannopoulos:2016dqw,Felegary:2016znh,Roy:2017mnz,Ghaffarnejad_2017,Lu:2019twd,Shabani_2019,Zucca:2019ohv,Giacomini:2020grc}. 

The idea of extra dimensions has also been an essential development for the unification of fundamental forces \cite{Bailin:1987jd}. The scalar fields from extra-dimensional theories such String/M theory are ubiquitous and play a crucial role in several phenomena such as the inflaton field \cite{Gasperini:1993hu,Damour:1995pd,Chamblin:1999ya}, DM \cite{Damour:1990tw}, dark radiation\cite{Svrcek:2006yi,Acharya:2015zfk} and cosmological constant \cite{Wetterich:1987fm,Shaposhnikov:2008xi}. In our previous work \cite{Pongkitivanichkul:2020txi}, we have proposed an extension of Einstein gravity using degrees of freedom coming from compactification of extra-dimensions, see \cite{Overduin:1998pn} for review and references therein. In particular, the toroidal 5-dimensional spacetime ($\mathbb{R}^4 \times S^1$) is the starting point and the 5-dimensional metric containing a scalar field and a vector field is assumed. This process of circular compactification is also known as Kaluza-Klein (KK) compactification. We also include 2 additional gauge fields in order to solve the anisotropy problem of the energy momentum tensor. In our previous paper, dynamical system analysis has been carried out and we demonstrated the potential to have both DM and DE phases of the universe. 

It was shown that including a matter field into the system can sometimes changes the behaviour of the dynamical system \cite{Hrycyna:2013hla, Hrycyna:2013yia}. In this paper, we would like to provide a more extensive study of our model. In particular, we are interested in the inclusion of matter field and radiation into the existing dynamical system in order to achieve the semi-realistic universe. The paper is organised as follows. First, we give a review on the model in section \ref{sec:The Model} where the additional matter and radiation field are also included. The full list of equation of motions will be given in section \ref{sec:EOM}. In section \ref{sec:Lower Mode Cases}, the range of compactification radius is chosen such that the number of scalar fields relevant to low energy physics will be finite and the numerical analysis can be done. The dynamical system analysis is performed in section \ref{sec:Dynamical System}. The full results will be given in section \ref{sec:Result} and the discussion of the results can be found in section \ref{sec:Discussion}. Finally, the conclusion is given in section \ref{sec:Conclusion and Outlook}.
\section{\label{sec:The Model}A Kaluza-Klein inspired Brans-Dicke Model}
In this section, we set up the crucial ingredients of the KK inspired BD model and use for studying the dynamical system analysis in the latter. First of all, the 5-dimensional action for KK gravity with free massive scalar field, $\Tilde{\eta}$ is given by \cite{Pongkitivanichkul:2020txi}
\begin{align}
    S=\int d^4 x dy \sqrt{-\Tilde{g}}\sum_{a=1}^2 \left( -\Tilde{\mathcal{R}}+\hat{\partial}_A \Tilde{\eta}^*\partial^A \Tilde{\eta}-M^2_{(5)} \Tilde{\eta}^* \Tilde{\eta} - \frac{1}{4}\phi^2\Tilde{F}^a_{AB}\Tilde{F}^{aAB} + V(\Tilde{A}^a_M) \right)
    ,
\end{align}
where $M_{(5)}$ is the mass of the scalar field $\tilde\eta$ and $\tilde{\mathcal{R}}$, $\phi$, $\tilde{F}_{AB}^a \equiv \partial_A\,A^a_B - \partial_B\,A^a_A$, $A_M^a$ are Ricci scalar, dilaton field, strength tensor and gauge field in 5 dimensions, respectively. In addition, we use natural unit system where $\sqrt{16\pi G} = 1$. The capital Latin indices represent the bulk dimensional spacetime indices, $A,B,C,\cdots = 0,1,2,3,5$ whereas the Latin indices stand for the triad gauge field indices, $a,b,c,\cdots = 1,2$. While $V( \tilde{A}_M^a)$ is the potential and it is composed of the the gauge field $\tilde{A}_M^a$ only and it will be specified in the later. Moreover, all variables with tilde symbol, $\tilde{~}$, are described the physical quantities in 5-dimensional spacetime. The metric tensor $\tilde g_{AB}$ of the bulk 5-dimensional spacetime is written by \cite{Overduin:1998pn},
\begin{eqnarray}
\widetilde{g}_{AB} = \left( 
\begin{array}{cc}
g_{\mu\nu} + \phi^2 A_\mu A_\nu & \phi^2 A_\mu
\\
\phi^2 A_\nu & \phi^2
\end{array}
\right).
\end{eqnarray}
It has been shown in Ref. \cite{Pongkitivanichkul:2020txi} that the gravitational action of the KK theory can be reduced to the 4-dimensional spacetime by mean of the compactification of the $y$ dimension. The full form of the KK theory in the 4 dimensions is written in the following form,
\begin{eqnarray}
S = S_{BD} + S_{KK} + S_G\,,
\label{full-action}
\end{eqnarray}
where $S_{BD}$, $S_{KK}$ and $S_G$ are the actions of the BD gravity, the 4 dimensional KK scalar field and the gauge fields, respectively. The explicit forms of three actions above are given by,
\begin{eqnarray}
S_{BD} &=& \int d^4 x \sqrt{-g}\,\phi\left[ -\mathcal{R} - \frac{1}{4}\phi^2F_{\mu\nu}F^{\mu\nu} - \frac{2}{3}\frac{\partial^{\mu}\phi \partial_{\mu}\phi}{\phi^2}\right],
\label{BD-action}
\\
S_{KK} &=& \int d^4 x \sqrt{-g}\,\phi \sum_{n=0}^{\infty}\left[ \partial_{\mu}\eta_n \partial^{\mu}\eta_n - M_{(5)}^2 \eta^2 + \left(\frac{1}{\phi^2} + A_{\nu}A^{\nu}\right)\frac{n^2}{R_k^2}\eta_n^2 \right],
\label{4d-scalar-KK}
\\   
S_G &=& \int d^4x \sqrt{-g}\,\phi \sum_{a=1}^{2}\sum_{n=0}^{\infty} \Bigg[ -\frac{1}{4}\phi^2\widetilde{F}^a_{n\mu\nu}\widetilde{F}^{a\mu\nu}_n + \left( V'' - \frac{\phi^2 n^2}{4 R_k^2} \right) \widetilde{A}^a_{\mu,n} \widetilde{A}^{\mu,a}_n   
\nonumber 
\\
&&\qquad\qquad\qquad\qquad\quad\quad   -\frac{1}{4}\phi^2 \partial_{\mu} \widetilde{A}_{5,n}^a \partial^{\mu}\widetilde{A}^{5,a}_n + V''_{(5)}\widetilde{A}^a_{5,n} \widetilde{A}^{5,a}_n \Bigg], 
\label{gauge-action}
\end{eqnarray}
where $V'' = \frac{\delta^2 V}{\delta A_{\mu}^2}$ and $V''_{(5)} = \frac{\delta^2 V}{\delta A_{5}^2}$. To obtain the actions in Eqs.(\ref{BD-action},\ref{4d-scalar-KK},\ref{gauge-action}), we have assumed that the $\eta_n$ and $\phi$ are real scalar fields. For all detail derivations of the actions, we refer to Ref.\cite{Pongkitivanichkul:2020txi} and we will not repeat them in this work. We close this section by summarising the main result in this model. We have shown that the BD gravity with scalar and gauge fields can be generated from the dimensional compactification of the 5-dimensional KK theory. The dilaton field is coupled to both Ricci scalar and the scalar field in 4-dimensional actions. In the following sections, we will use the KK inspired BD model in the presence of the barotropic fluid matter to demonstrate that the dilaton, scalar and gauge fields might play the role of DM and DE.

\section{\label{sec:EOM}Equation of Motions}
In this section, we compute the equations of motion of the KK inspired BD model and we will employ the results to investigate the dynamics of the universe in this model as a candidate of DM and DE. With the flat FLRW metric, $ds^2 = dt^2 - a^2(t)(dx^2+dy^2+dz^2)$, we firstly obtain the equation of motion for the dilaton field, $\phi$ from the Euler-Lagrange equation as
\begin{align}
       \mathcal{R} =&~ \frac{4}{3\sqrt{-g}\phi} \partial^{\mu}(\sqrt{-g}\partial_{\mu}\phi) - \frac{3}{4}\phi^2 F_{\mu\nu} F^{\mu\nu} - \frac{2}{3\phi^2}\partial^{\mu}\phi\partial_{\mu}\phi \nonumber\\
        & + \sum_{n=0}^{\infty}\left[ \partial_{\mu}\eta_n \partial^{\mu}\eta_n - M_{(5)}^2 \eta^2 + \left(\frac{1}{\phi^2} + A_{\nu}A^{\nu}\right)\frac{n^2}{R_k^2}\eta_n^2 \right] \nonumber\\
        & +  \sum_{a=1}^{2}\sum_{n=0}^{\infty} \left( -\frac{3}{4}\phi^2\widetilde{F}^a_{n\mu\nu}\widetilde{F}^{a\mu\nu}_n + \left( V'' - \frac{3}{4}\frac{\phi^2 n^2}{ R_k^2} \right) \widetilde{A}^a_{\mu,n} \widetilde{A}^{\mu,a}_n \right) 
        .
\end{align}
According to the action in Eq.(\ref{full-action}), the matter Lagrangian together with the barotropic fluid can defined as 
\begin{eqnarray}
\label{eq:Lmatter}
     \mathcal{L}_{matter} &=& -\frac{1}{4}\phi^{3}F^{2}-\frac{2}{3}\frac{\partial_{\mu}\phi\partial^{\mu}\phi}{\phi} 
     \nonumber\\
     && +\, \phi \left[ \partial_{\mu}\eta_n \partial^{\mu}\eta_n - M_{(5)}^2 \eta^2 + \left(\frac{1}{\phi^2} + A_{\nu}A^{\nu}\right)\frac{n^2}{R_k^2}\eta_n^2 \right] 
          + \mathcal{L}_{fluid} 
        ,
\end{eqnarray} where $\mathcal{L}_{fluid}$ represents other fields (particles) which could be existing in the 4 dimensions.
The energy-momentum tensor can be determined by
\begin{align}
\label{eq:T}
    T^{\mu\nu}=-\frac{2}{\sqrt{-g}}\frac{\delta(\sqrt{-g}\mathcal{L}_{matter})}{\delta g^{\mu\nu}}.
\end{align}
Substituting Eq.(\ref{eq:Lmatter}) into Eq.(\ref{eq:T}), we find
\begin{align}
     T_{\mu\nu}=&~\phi^{3}F_{\mu\alpha}F_{\nu}\!^{\alpha}-\frac{g_{\mu\nu}}{4}\phi^3 F^2 - \frac{g_{\mu\nu}}{4}\phi^3 \sum_{a=1}^2 \Tilde{F}_{\mu\nu,a}\Tilde{F}^{\mu\nu}_a + \frac{2}{3}\frac{\partial_{\mu}\phi\partial_{\nu}\phi}{\phi} - \phi \sum_{n=0}^{\infty} \partial_{\mu}\eta_{n}\partial_{\nu}\eta_{n} \nonumber\\
        & - 2 \phi A_{\mu}A_{\nu}\sum_{n=0}^{\infty} \frac{n^2\eta_n^2}{R_k^2} + g_{\mu\nu} \phi \sum_{n=0}^{\infty}\left[\left(\frac{1}{\phi^2} +  A_{\nu}A^{\nu}\right)\frac{n^2}{R_k^2}\eta_n^2 - M_{(5)}^2 \eta_n^2\right] \nonumber \\
        & + \phi^3 \sum_{n=0}^{\infty} \widetilde{F}^a_{\mu\alpha,n} \widetilde{F}^{a,n}_{\nu} - \phi \sum_{n=0}^{\infty} \left( V'' - \frac{n^2 \phi^2}{4 R_k^2} \right) (2 \widetilde{A}^a_{\mu,n} \widetilde{A}^{a}_{\nu,n} - g_{\mu\nu} \widetilde{A}^a_{\rho,n} \widetilde{A}^{\rho,a}_{n}) + \widetilde{T}_{\mu\nu}\,. 
        \label{energy-momentum-tensor}
\end{align}
The barotropic fluid energy-momentum, $\widetilde{T}_{\mu\nu}$ is given by
\begin{eqnarray}
\widetilde{T}_{\mu\nu} = (\rho_m + \rho_r)u_\mu u_\nu + p_r g_{\mu\nu}\,,
\end{eqnarray}
where $u_\mu$ is the usual comoving 4-velocity, whereas $\rho_m$, $\rho_r$ and $p_r$ are the energy density of dust matter, radiation and pressure of radiation, respectively. In order to fulfill the isotropy of the energy momentum tensor, the zero mode configuration of each species $(a =1,2)$ needs to satisfy
\begin{align}
    A^{\mu} =& (0,A,0,0)\\
   \widetilde{A}^{\mu,1}_{0} \equiv \widetilde{A}^{\mu,1}  =& (0,0,A,0)\\
   \widetilde{A}^{\mu,2}_{0} \equiv \widetilde{A}^{\mu,2} =& (0,0,0,A).
\end{align}
This specific configuration is also known as the Cosmic Triad \cite{ArmendarizPicon:2004pm,Golovnev:2008cf,Maleknejad:2012fw}.
In addition, the diagonal spatial components of the energy-momentum tensor in Eq.(\ref{energy-momentum-tensor})
require that 
\begin{equation}
   V'' = \sum_{n=0}^{\infty} \frac{n^2\eta_n^2}{R_k^2} \equiv M_A^2.
\end{equation}
The Einstein field equation of the KK inspired BD gravity takes the form of
\begin{equation}
     \phi(\mathcal{R}_{\mu\nu}-\frac{1}{2}g_{\mu\nu}\mathcal{R}) + g_{\mu\nu}\nabla_{\sigma}\nabla^{\sigma}\phi - \nabla_{\mu}\nabla_{\nu}\phi = T_{\mu\nu}.
     \label{EFE}
\end{equation}
Considering the $tt$ component, the Einstein tensor reads 
\begin{equation}
G_{tt}=\frac{\dot{a}^2}{a^2},
\end{equation}
with the fact that
\begin{align}
    \widetilde{F}^{t\sigma,1} \widetilde{F}^{t,1}\!_{\sigma} &= \widetilde{F}^{t\sigma,2}\widetilde{F}^{t,2}\!_{\sigma} = F^{t\sigma}F^{t}\!_{\sigma} = g_{ij}F^{ti}F^{tj} = -a^2\left(\dot{A} + 2 H A\right)^2 \\
    \widetilde{F}^{1}_{\mu\nu} \widetilde{F}^{\mu\nu,1} &= \widetilde{F}^{2}_{\mu\nu} \widetilde{F}^{\mu\nu,2} =  F_{\mu\nu}F^{\mu\nu} = F_{ti}F^{ti} + F_{it}F^{it} = -2a^2\left(\dot{A} + 2 H A\right)^2\,.
\end{align}
Then the Friedmann equation in the KK inspired BD model can be written as
\begin{eqnarray}
        \left(\frac{\dot{a}}{a}\right)^{2} &=& -\frac{3\phi^2a^2}{2}\left(\dot{A} + 2 H A\right)^2 +\frac{2\dot{\phi}^2}{3\phi^2} 
        \nonumber\\
        && -\, \sum_{n=0}^{\infty}\dot{\eta}_n^2  + \sum_{n=0}^{\infty} \left[ \left(\frac{1}{\phi^2} -3 a^2 A^2\right)\frac{n^2}{R_k^2} - M_{(5)}^2 \right]\eta_n^2 + \frac{\rho_m + \rho_r}{\phi} .
\end{eqnarray}
Recalling relation $\dot{H}=\frac{\Ddot{a}}{a}-\frac{\dot{a}^2}{a^2}$, the Reychaudhuri equation can be determined from the spatial components of the field equation in Eq.(\ref{EFE}) and it reads
\begin{align}
    - 3\ddot{\phi} =& ~ 3\phi H^2 + 6\phi \left(\dot{H} + H^2\right) + 9 H \dot{\phi} \nonumber \\
     -& \frac{3 a^2\phi^3}{2} \left(\dot{A} + 2 H A\right)^2 - 6 \phi a^2 A^2 \sum_{n=0}^{\infty} \frac{n^2\eta_n^2}{R_k^2} \nonumber\\
    -& 3\phi \sum_{n=0}^{\infty}\left[\left(\frac{1}{\phi^2} -3 a^2 A^2\right)\frac{n^2}{R_k^2}\eta_n^2 - M_{(5)}^2 \eta_n^2\right] + a^2\textcolor{black}{p_r}. \label{eq:EFEij}
\end{align}
In this section, we have prepared relevant equations of motion in the KK inspired BD gravity model to study the dynamics of the universe in the later. Since, the scalar field $\eta_n$ represents the infinite summation of the KK exited states and leads to complicated physical quantities in the model. Therefore we will consider only the lower mode of the KK excited states in the next section. 
\section{\label{sec:Lower Mode Cases}Lower Mode Cases}
It is well known that the KK scalar fields are composed of the infinite tower mode. In particular, a higher mode with a momentum larger than the reduced Plank scale will be physically irrelevant to the observable universe. Therefore, we can neglect such higher modes and the remaining modes of the KK excited field should satisfy the momentum condition:
\begin{equation}
n < R_k\,.
\end{equation}
Therefore, we focus to study the KK inspired BD model for, $1 < R_k < 2$. In this case, only zero mode and the first excited mode involve in the system. For simplicity and convenient to study the dynamics of the universe by using the dynamical system analysis with the dimensionless parameters, we can simply drop the radiation matter term in this work. Moreover, we found that the inclusion of radiation part would lead to an unnecessary introduction of free parameters. The Friedmann equation for the lower mode of the model reads,
\begin{eqnarray}
H^{2} = -\frac{3 \phi^2a^2}{2}\mathcal{A}_c^2 +\frac{2\dot{\phi}^2}{3\phi^2} - \dot{\eta}_0^2 - \dot{\eta}_1^2 - M_{(5)}^2 \eta_0^2 - M_{(5)}^2 \eta_1^2  +  \left(\frac{1}{\phi^2} - 3 a^2 A^2\right) \frac{\eta_1^2}{R_k^2} + \frac{\rho_m}{\phi}. \label{eq:Friedman2}
\end{eqnarray}
Noted that $\eta_0$ and $\eta_1$ scalar fields are the zero mode and the first excited mode, respectively. The equation of motion for the dilation field, $\phi$ becomes
\begin{align}
    \ddot{\phi} + 3 H \dot{\phi} =& -\frac{9}{2}\phi\left(\dot{H} + 2H^2\right) - \frac{27}{8}\phi^3a^2 \mathcal{A}_c^2 + \frac{\dot{\phi}^2}{2\phi} - \frac{3}{4}\phi\dot{\eta}_0^2 - \frac{3}{4}\phi\dot{\eta}_1^2 \nonumber \\
& + \frac{3}{4} \phi \left( \frac{1}{\phi^2} + 3 a^2 A^2 \right)\frac{\eta_1^2}{R_k^2} + \frac{3}{4}\phi M_{(5)}^2\eta_0^2 + \frac{3}{4}\phi M_{(5)}^2\eta_1^2 \label{eq:seteomphi}.
\end{align}
The pressure can be written as $p_i=w_i\rho_i$ where $w_m =0$. Solving Eq.(\ref{eq:EFEij}) and Eq.(\ref{eq:seteomphi}) gives
\begin{align}
    \dot{H} =& \frac{13  a^2 A^2 \eta_1^2}{10 R_k^2}-\frac{31}{20} a^2 \phi^2 \mathcal{A}_c^2-\frac{3}{10}  \dot{\eta}_1^2+\frac{7}{10} M_{(5)}^2  \eta_1^2-\frac{  \eta_1^2}{10 R_k^2 \phi^2}\nonumber \\
    & -\frac{12 H^2}{5}-\frac{3}{10}  \dot{\eta}_0^2+\frac{7}{10} M_{(5)}^2  \eta_0^2+\frac{\dot{\phi}^2}{5 \phi^2} ,
\end{align}
and 
\begin{align}
    \ddot{\phi} =& -\frac{18  a^2 A^2 \eta_1^2 \phi}{5 R_k^2}+\frac{18}{5} a^2 \phi^3 \mathcal{A}_c^2+\frac{3}{5}   \phi \dot{\eta}_1^2-\frac{12}{5} M_{(5)}^2  \eta_1^2 \phi + \frac{6   \eta_1^2}{5 R_k^2 \phi}\nonumber \\
    & -3 H \dot{\phi} + \frac{9}{5} H^2 \phi + \frac{3}{5}   \phi \dot{\eta}_0^2-\frac{12}{5} M_{(5)}^2  \eta_0^2 \phi - \frac{2 \dot{\phi}^2}{5 \phi},
\end{align}
where we define $\mathcal{A}_c = \dot{A} + 2 H A$\,.

The specific form of equations and fields allow us to analyse the system both analytically and numerically. In the next section, we will use the results to study the dynamical system of the model.
\section{\label{sec:Dynamical System}Dynamical System}
Now, we arrive at the crucial part of this work that is to study the dynamics of the universe. The main purpose of this work is to demonstrate the existence of DM and DE in the KK inspired BD gravity with the inclusion of the barotropic fluid. Since the dynamical system is widely used and succeed to qualitatively analyse the cosmological models \cite{Bahamonde:2017ize}, we will apply this method to the model. For simplicity, we define the dimensionless parameters to construct the autonomous system of the equations,

\begin{eqnarray}
&& X_1 = \frac{\sqrt{3}a}{H}\mathcal{A}_c,\quad X_2 = \sqrt{\frac{2}{3}}\frac{\dot{\phi}}{H\phi},\quad X_3 = \frac{\dot{\eta}_0}{H},\quad X_4 = \frac{\dot{\eta}_1}{H},\quad
X_5 = \frac{1}{\phi},\nonumber \\ 
&& X_6 = \sqrt{3} a A,\quad X_7 = \frac{\eta_1}{H R_k},\quad X_8 = \frac{\eta_0}{H R_k}, \quad X_9=\sqrt{\frac{\rho_m}{H^2}}.
\end{eqnarray}
We can use the Friedmann equation as a constraint equation of the the dimensionless parameters, i.e., 
\begin{equation}
1=-\frac{X_1^2}{2 X_5^2}+X_2^2-X_3^2-X_4^2-X_6^2 X_7^2+X_5 \left(X_5 X_7^2+X_9^2\right)-M_{(5)}^2 R_k^2. \left(X_7^2+X_8^2\right)\label{eq:cont}
\end{equation}
Noted that the additional free parameters of the system are defined by
\begin{equation}
    \lambda \equiv M_{(5)}R_k \quad\text{and}\quad \mu \equiv \frac{1}{H R_k}.
\end{equation}
Then the conformal derivative of the defined parameters are derived as following
\begin{align}
\frac{1}{H}\frac{dX_1}{dt}=&\frac{31 X_1^3}{60 X_5^2}-\frac{3 X_1 X_2^2}{10}+\frac{3 X_1 X_3^2}{10}+\frac{3 X_1 X_4^2}{10}+\frac{1}{10} X_1 X_5^2 X_7^2-\frac{13}{30} X_1 X_6^2 X_7^2\,,\nonumber \\
&-\frac{7}{10} \lambda ^2 X_1 X_7^2-\frac{7}{10} \lambda ^2 X_1 X_8^2+\frac{17 X_1}{5}\label{eq:1}\\
\frac{1}{H}\frac{dX_2}{dt}=&\frac{31 X_1^3}{60 X_5^2}-\frac{3 X_1 X_2^2}{10}+\frac{3 X_1 X_3^2}{10}+\frac{3 X_1 X_4^2}{10}+\frac{1}{10} X_1 X_5^2 X_7^2-\frac{13}{30} X_1 X_6^2 X_7^2\nonumber \\
&-\frac{7}{10} \lambda ^2 X_1 X_7^2-\frac{7}{10} \lambda ^2 X_1 X_8^2+\frac{17 X_1}{5}\,,\label{eq:2}\\
\frac{1}{H}\frac{dX_3}{dt}=&\frac{31 X_1^2 X_3}{60 X_5^2}-\frac{3 X_2^2 X_3}{10}-\sqrt{\frac{3}{2}} X_2 X_3+\frac{3 X_3^3}{10}+\frac{3 X_3 X_4^2}{10}+\frac{1}{10} X_3 X_5^2 X_7^2\nonumber \\
&-\frac{13}{30} X_3 X_6^2 X_7^2-\frac{7}{10} \lambda ^2 X_3 X_7^2-\frac{7}{10} \lambda ^2 X_3 X_8^2-\frac{3 X_3}{5}-\lambda ^2 \mu  X_8\,,\label{eq:3}\\
\frac{1}{H}\frac{dX_4}{dt}=&\frac{31 X_1^2 X_4}{60 X_5^2}-\frac{3 X_2^2 X_4}{10}-\sqrt{\frac{3}{2}} X_2 X_4+\frac{3 X_3^2 X_4}{10}+\frac{3 X_4^3}{10}+\frac{1}{10} X_4 X_5^2 X_7^2-\lambda ^2 \mu  X_7\nonumber \\
&-\frac{13}{30} X_4 X_6^2 X_7^2-\frac{7}{10} \lambda ^2 X_4 X_7^2-\frac{7}{10} \lambda ^2 X_4 X_8^2-\frac{3 X_4}{5}+\mu  X_5^2 X_7-\mu  X_6^2 X_7\,,\label{eq:4}\\
\frac{1}{H}\frac{dX_5}{dt}=&-\sqrt{\frac{3}{2}} X_2 X_5\,,\label{eq:5}\\
\frac{1}{H}\frac{dX_6}{dt}=&X_1-X_6\,,\label{eq:6}\\
\frac{1}{H}\frac{dX_7}{dt}=&\frac{31 X_1^2 X_7}{60 X_5^2}-\frac{3 X_2^2 X_7}{10}+\frac{3 X_3^2 X_7}{10}+\frac{3 X_4^2 X_7}{10}+\mu  X_4+\frac{X_5^2 X_7^3}{10}-\frac{13 X_6^2 X_7^3}{30}\,,\nonumber \\
&+\frac{12 X_7}{5}-\frac{7 \lambda ^2 X_7^3}{10}-\frac{7}{10} \lambda ^2 X_7 X_8^2\,,\label{eq:7}\\
\frac{1}{H}\frac{dX_8}{dt}=&\frac{31 X_1^2 X_8}{60 X_5^2}-\frac{3 X_2^2 X_8}{10}+\frac{3 X_3^2 X_8}{10}+\mu  X_3+\frac{3 X_4^2 X_8}{10}+\frac{1}{10} X_5^2 X_7^2 X_8\nonumber\\
&-\frac{13}{30} X_6^2 X_7^2 X_8-\frac{7}{10} \lambda ^2 X_7^2 X_8-\frac{1}{10} 7 \lambda ^2 X_8^3+\frac{12 X_8}{5}\,,\label{eq:8}\\
\frac{1}{H}\frac{dX_9}{dt}=&\frac{31 X_1^2 X_9}{60 X_5^2}-\frac{3 X_2^2 X_9}{10}+\frac{3 X_3^2 X_9}{10}+\frac{3 X_4^2 X_9}{10}+\frac{1}{10} X_5^2 X_7^2 X_9-\frac{13}{30} X_6^2 X_7^2 X_9\nonumber\\
&-\frac{7}{10} \lambda^2 X_7^2 X_9-\frac{7}{10} \lambda^2 X_8^2 X_9+\frac{9 X_9}{10}\,,\label{eq:9} \\
\frac{1}{H}\frac{d\mu}{dt}=&\frac{12 \mu }{5}+\frac{31 \mu  X_1^2}{60 X_5^2}-\frac{3 \mu X_2^2}{10}+\frac{3 \mu X_3^2}{10}+\frac{3 \mu X_4^2}{10}+\frac{1}{10} \mu X_5^2 X_7^2-\frac{13}{30} \mu X_6^2 X_7^2\nonumber \\
&-\frac{7}{10} \lambda ^2 \mu X_7^2-\frac{7}{10} \lambda ^2 \mu X_8^2. \label{eq:mu}
\end{align}
We obtain a critical point of the dynamical system by setting all derivative equations from Eqs.(\ref{eq:1}-\ref{eq:9}) equal to zero with constraint Eq.(\ref{eq:cont}). The stability of a critical point can be analysed from its eigenvalues of the matrix
\begin{equation}
    M_{ij}=\frac{d}{dX_i}\left( \frac{1}{H} \frac{dX_j}{dt} \right) \label{eq:eigenv}.
\end{equation}
If all eigenvalues are negative, the critical point is stable and if any is positive, the critical point is unstable. The energy density of each specie can be written as

\begin{align}
    &\Omega_{A} = -\frac{X_1^2}{2X_5^2} -     X_6^2 X_7^2,\quad \Omega_{\phi} = X_2^2 + X_5^2 X_7^2,\quad
    \Omega_{\eta_0} = - \left(X_3^2 + \lambda^2 X_8^2\right),\nonumber\\
    &\Omega_{\eta_1} = - \left(X_4^2 + \lambda^2 X_7^2\right),\quad
    \Omega_{\rho_m} = X_5X_9^2.
    \label{density}
\end{align}
The effective equation of state is given by
\begin{equation}
    w_\text{eff}=-1-\frac{2\dot{H}}{3H^2}.
    \label{w-eff}
\end{equation}
We use the above equation to identify the phase of the universe, where $w_\text{eff}<-1/3$ and $w_\text{eff}>-1/3$ represent the acceleration and deceleration expansion of the universe, respectively. While $w_\text{eff}=0$ represents the matter dominated phase of the universe, $w_\text{eff}<-1$ corresponds to the phantom DE. In addition, one may find the solutions of the scale factor by integrating the Eq.(\ref{w-eff}).
\begin{table}[b]
\centering
\begin{tabular}{|c|c|c|c|c|c|}
\hline
System & Non-vanishing fields & Existence & Stability & $w_\text{eff}$ & Possible phase \\ \hline
\hyperlink{FA1}{$\mathcal{F}_1^{(A)}$} & $\phi, \eta_1,\rho_m$ &$\lambda >0 \quad\land\quad \mu >\frac{3}{2 \lambda }$ & \textcolor{blue}{Stable} & 0 & DM \\ \hline
\hyperlink{FA2}{$\mathcal{F}_2^{(A)}$} & $\phi, \eta_1,\rho_m$ &$\lambda >0 \quad\land\quad \mu >\frac{3}{2 \sqrt{2} \lambda }$ & \textcolor{blue}{Stable} & 0 & DM \\ \hline
\hyperlink{FB1}{$\mathcal{F}_1^{(B)}$} & $\phi, \eta_1, A, \rho_m$ &$\lambda >0 \quad\land\quad \mu >0$ & \textcolor{red}{\it{Saddle}} & 0 & DM \\ \hline
\hyperlink{FB2}{$\mathcal{F}_2^{(B)}$} & $\phi, \eta_1, A, \rho_m$ &$\lambda >0 \quad\land\quad \mu >0$ & \textcolor{blue}{Stable} or \textcolor{red}{\it{Saddle}} & $<-1$ & Phantom DE \\ \hline
\hyperlink{FB3}{$\mathcal{F}_3^{(B)}$} & $\phi, \eta_1, A, \rho_m$ &$\lambda >0\quad\land\quad\mu >\frac{1}{\lambda }\sqrt{\frac{57}{10}}$ & \textcolor{red}{\it{Saddle}} & $-5/3$ & Phantom DE \\ \hline
\hyperlink{FC1}{$\mathcal{F}_1^{(C)}$} & $\phi, \eta_0, \eta_1, \rho_m$ &$\lambda >0 \quad\land\quad \mu >\frac{3}{2 \lambda }$ & \textcolor{blue}{Stable} & $0$ & DM \\ \hline
\hyperlink{FD1}{$\mathcal{F}_1^{(D)}$} & $\phi, \eta_0, \eta_1, A, \rho_m$ &$\lambda >0 \quad\land\quad \mu >0$ & \textcolor{red}{\it{Saddle}} & $0$ & DM \\ \hline
\hyperlink{FD2}{$\mathcal{F}_2^{(D)}$} & $\phi, \eta_0, \eta_1, A, \rho_m$ &$\lambda >0\quad\land\quad \mu >\frac{1}{\lambda }\sqrt{\frac{57}{10}}$ & \textcolor{red}{\it{Saddle}} & $-5/3$ & Phantom DE \\ \hline
\hyperlink{FBbar1}{$\bar{\mathcal{F}}_1^{(B)}$} & $\phi, \eta_1, A, \rho_m$ &$\lambda >0\quad\land\quad \mu >0$ & \textcolor{blue}{Stable} & $-1$ & DE \\ \hline
\hyperlink{FBbar2}{$\bar{\mathcal{F}}_2^{(B)}$} & $\phi, \eta_1, A, \rho_m$ &$\lambda >0\quad\land\quad \mu >0$ & \textcolor{red}{\it{Saddle}} & $-1$ & DE \\ \hline
\hyperlink{FBbar3}{$\bar{\mathcal{F}}_3^{(B)}$} & $\phi, \eta_1, A, \rho_m$ &$\lambda >0\quad\land\quad \mu >0$ & \textcolor{red}{\it{Saddle}} & $-1$ & DE \\ \hline
\hyperlink{FBbar4}{$\bar{\mathcal{F}}_4^{(B)}$} & $\phi, \eta_1, A, \rho_m$ &$\lambda >0\quad\land\quad \mu >0$ & \textcolor{red}{\it{Saddle}} & $-1/3$ & Critical DE \\ \hline
\hyperlink{FBbar5}{$\bar{\mathcal{F}}_5^{(B)}$} & $\phi, \eta_1, A, \rho_m$ &$\lambda >0\quad\land\quad \mu >0$ & \textcolor{blue}{Stable} & $-5/3$ & Phantom DE \\ \hline
\hyperlink{FCbar1}{$\bar{\mathcal{F}}_1^{(C)}$} & $\phi, \eta_0, \eta_1, \rho_m$ &$\lambda >0\quad\land\quad \mu >0$ & \textcolor{blue}{Stable} & $-1$ & DE \\ \hline
\hyperlink{FCbar2}{$\bar{\mathcal{F}}_2^{(C)}$} & $\phi, \eta_0, \eta_1, \rho_m$ &$\lambda >0\quad\land\quad \mu >0$ & \textcolor{red}{\it{Saddle}} & $-1/3$ & Critical DE \\ \hline
\hyperlink{FCbar3}{$\bar{\mathcal{F}}_3^{(C)}$} & $\phi, \eta_0, \eta_1, \rho_m$ &$\lambda >0\quad\land\quad \mu >0$ & \textcolor{blue}{Stable} & $-1$ & DE \\ \hline
\hyperlink{FDbar1}{$\bar{\mathcal{F}}_1^{(D)}$} & $\phi, \eta_0, \eta_1, A, \rho_m$ &$\lambda >0\quad\land\quad \mu >0$ & \textcolor{blue}{Stable} & $-1$ & DE \\ \hline
\hyperlink{FDbar2}{$\bar{\mathcal{F}}_2^{(D)}$} & $\phi, \eta_0, \eta_1, A, \rho_m$ &$\lambda >0\quad\land\quad \mu >0$ & \textcolor{red}{\it{Saddle}} & $-1$ & DE \\ \hline
\hyperlink{FDbar3}{$\bar{\mathcal{F}}_3^{(D)}$} & $\phi, \eta_0, \eta_1, A, \rho_m$ &$\lambda >0\quad\land\quad \mu >0$ & \textcolor{red}{\it{Saddle}} & $-1/3$ & Critical DE \\ \hline
\hyperlink{FDbar4}{$\bar{\mathcal{F}}_4^{(D)}$} & $\phi, \eta_0, \eta_1, A, \rho_m$ &$\lambda >0\quad\land\quad \mu >0$ & \textcolor{blue}{Stable} & $-5/3$ & Phantom DE \\ \hline
\end{tabular}

\caption{The table summarises non-vanishing fields, effective equation of state and possible phases of the universe for each system. Noted that $\bar{\mathcal{F}}_i$ are the systems with the slow-roll approximation, $\dot{\eta_i}=0$.} \label{tab:tab1}
\end{table}
\section{\label{sec:Result}Results}
Since there are 9 main dimensionless parameters, $X_1-X_9$ and 1 extra parameter, $\mu$. It is more convenient to categorise the dynamical system into many cases. First, let's consider the case that the 5-dimensional mass, $M_{(5)}$ does not exist in the universe. Therefore, the $X_8$ is decoupled from the Friedmann equation. We analyse with assumption, $X_3=X_8=0$, and found that there is no real solution in this case. 
Another case is the assumption that $M_{(5)}\neq0$. We will consider this in various sets of the non-vanishing fields in the dynamical system. All possible real solutions are summarised in Tab.(\ref{tab:tab1}). Moreover, the slow-roll approximation of the $\dot\eta_{0}$ and $\dot\eta_{1}$ is taken into account for some of these solutions.

\subsection{\label{sec:M5!=0}$M_{(5)}\neq0$ case}
In the case of $M_{(5)}\neq0$, there are many solutions of the critical point and they are written in terms of the parameters $\lambda$ and $\mu$. In the analysis, the Eqs.(\ref{eq:cont},\ref{eq:2},\ref{eq:5}) prevent $X_2=X_5=0$ which mean that $\phi$ cannot be decoupled from the system. The critical points of the relevant non-vanishing fields in the autonomous system are classified below. 


\subsubsection{$\mathcal{F}_i^{(A)}$ : $(\phi,\eta_1,\rho_m)$-system (A) }
First, we consider $(\phi, \eta_1, \rho_m)$-system which leads to $X_1=X_3=X_6=X_8=0$, with the following constraint equation,
\begin{equation}
    1=X_2^2 - X_4^2 +X_5^2X_7^2 + X_5X_9^2-\lambda^2X_7^2. \label{eq:contA}
\end{equation}
In order to find the critical points, we use the constraint equation, Eq.(\ref{eq:contA}), to reduce the autonomous system by replacing a parameter for the remaining differential equation.\par
$\bullet$ \hypertarget{FA1}{$\mathcal{F}_1^{(A)}$}, we eliminate $X_2$ by using the constraint equation in Eq.(\ref{eq:contA}). The critical point is given by
\allowdisplaybreaks
\begin{align}
X_4=& -\frac{3}{\sqrt{\frac{8 \lambda ^2 \mu ^2}{3}-2}},
\quad X_5= \frac{\sqrt{4 \lambda ^2 \mu ^2-9}}{2 \mu },
\nonumber\\
X_7=& \frac{\sqrt{6} \mu }{\sqrt{4 \lambda ^2 \mu ^2-3}},
\quad X_9= -2\sqrt{2} \sqrt{\frac{\mu  \left(4 \lambda ^4 \mu ^4+15 \lambda ^2 \mu ^2-54\right)}{\left(4 \lambda ^2 \mu ^2-9\right)^{3/2} \left(4 \lambda ^2 \mu ^2-3\right)}}.
\end{align}
The effective equation of state of this system then simply reduces to
\begin{equation}
    w_\text{eff}=0.
\end{equation}
The critical point exists for all positive $\lambda$ and $\mu$ with $\lambda\mu > \frac{3}{2}$. After using the definition of stability matrix in Eq.(\ref{eq:eigenv}), the real part of eigenvalues is read
\begin{equation}
\left( 0, 0, 0, -\frac{9}{5} \right).
\end{equation}
This means that this critical point is a stable point. Then, the $\mathcal{F}_1^{(A)}$ can be represented as the late time DM dominated phase. The density parameters are given by
\begin{align}
\Omega_{\phi }=\frac{3 \left(4 \lambda ^2 \mu ^2-9\right)}{8 \lambda^2 \mu ^2-6},\quad
\Omega_{\eta _1}=\frac{3 \left(4 \lambda ^2 \mu ^2+9\right)}{6-8 \lambda ^2 \mu ^2},\quad
\Omega_{\rho _m}=\frac{4 \left(\lambda ^2 \mu ^2+6\right)}{4 \lambda ^2 \mu ^2-3}.
\end{align}
\begin{figure}[H]%
    \centering
    \subfloat[\centering Stability]{{\includegraphics[height=5.5cm]{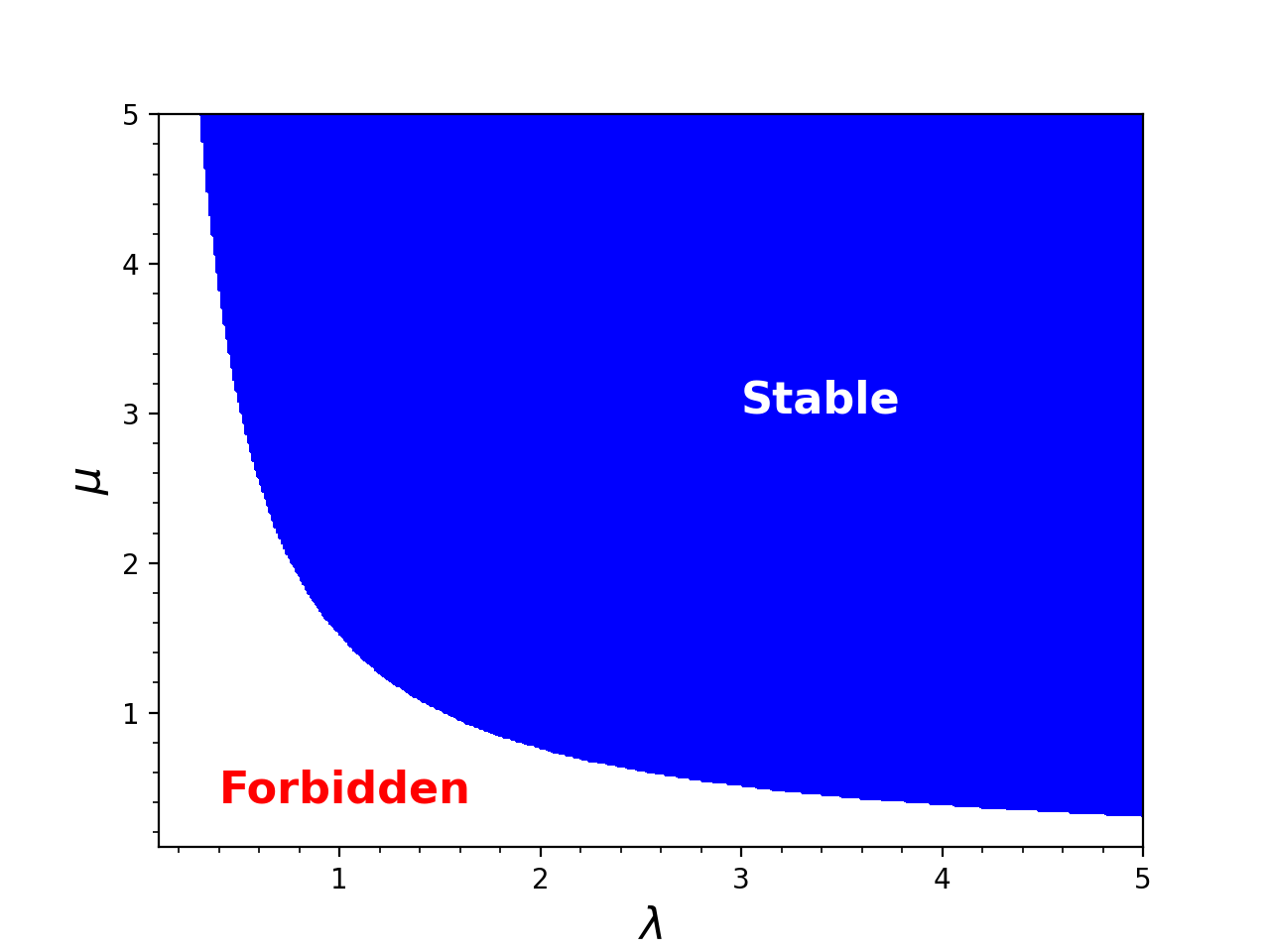}}}%
    \subfloat[\centering Density parameters]{{\includegraphics[height=5.5cm]{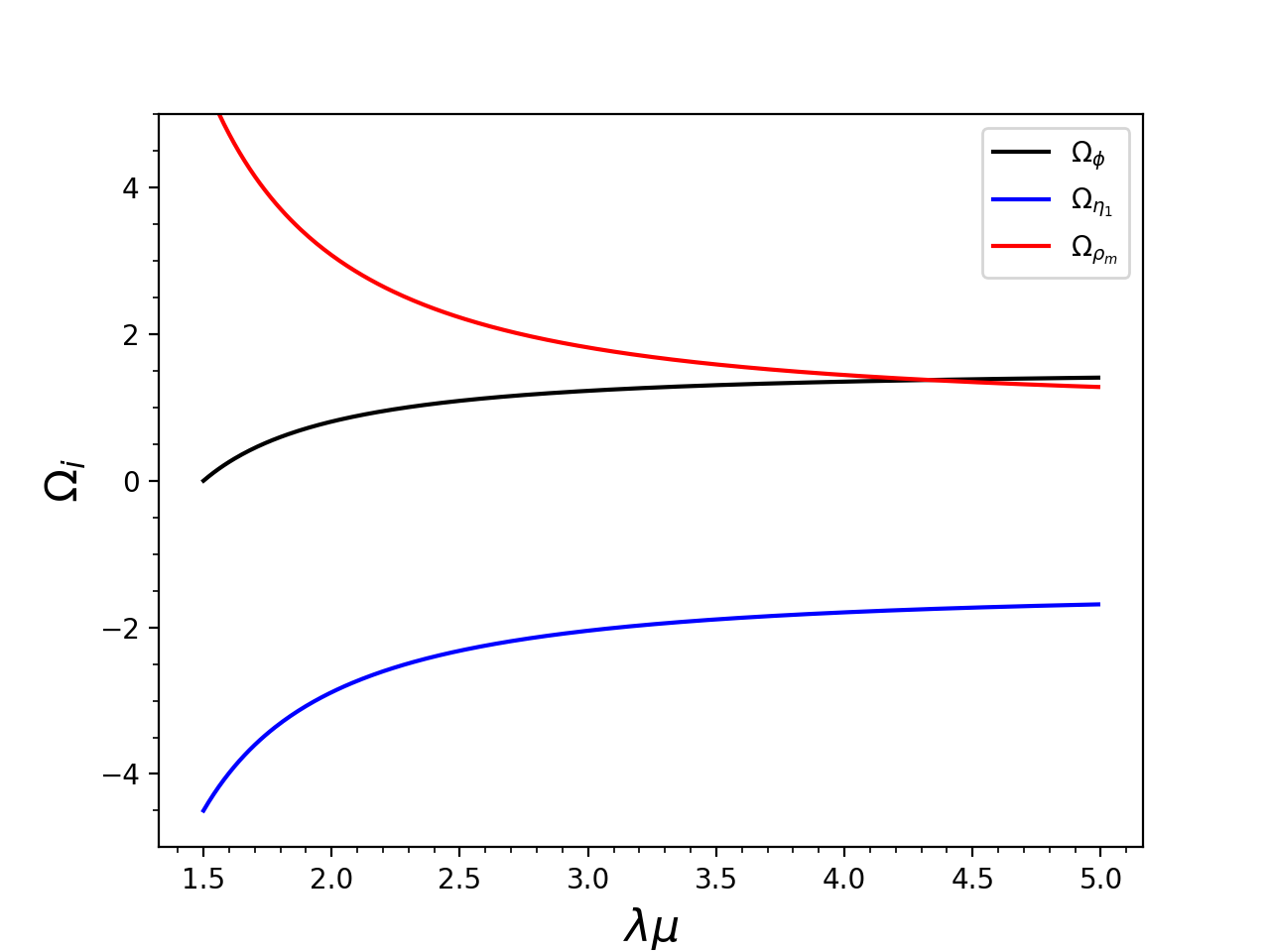}}}%
    \caption{On the left panel, the graph (a) shows the stability of the $\mathcal{F}_1^{(A)}$ critical point. The stable critical points lie in the blue shaded region while non-shaded region is forbidden for critical points. On the right panel, the graph (b) illustrates the relation between density parameter of non-vanishing fields and free parameters, i.e., $\lambda$ and $\mu$. The black, blue and red line in graph (b) indicates $\Omega_\phi$, $\Omega_{\eta_1}$ and $\Omega_{\rho_m}$, respectively.}
    \label{fig:plotFA1}%
\end{figure}

\begin{figure}[H]%
    \centering
    \subfloat[\centering Stability]{{\includegraphics[height=5.5cm]{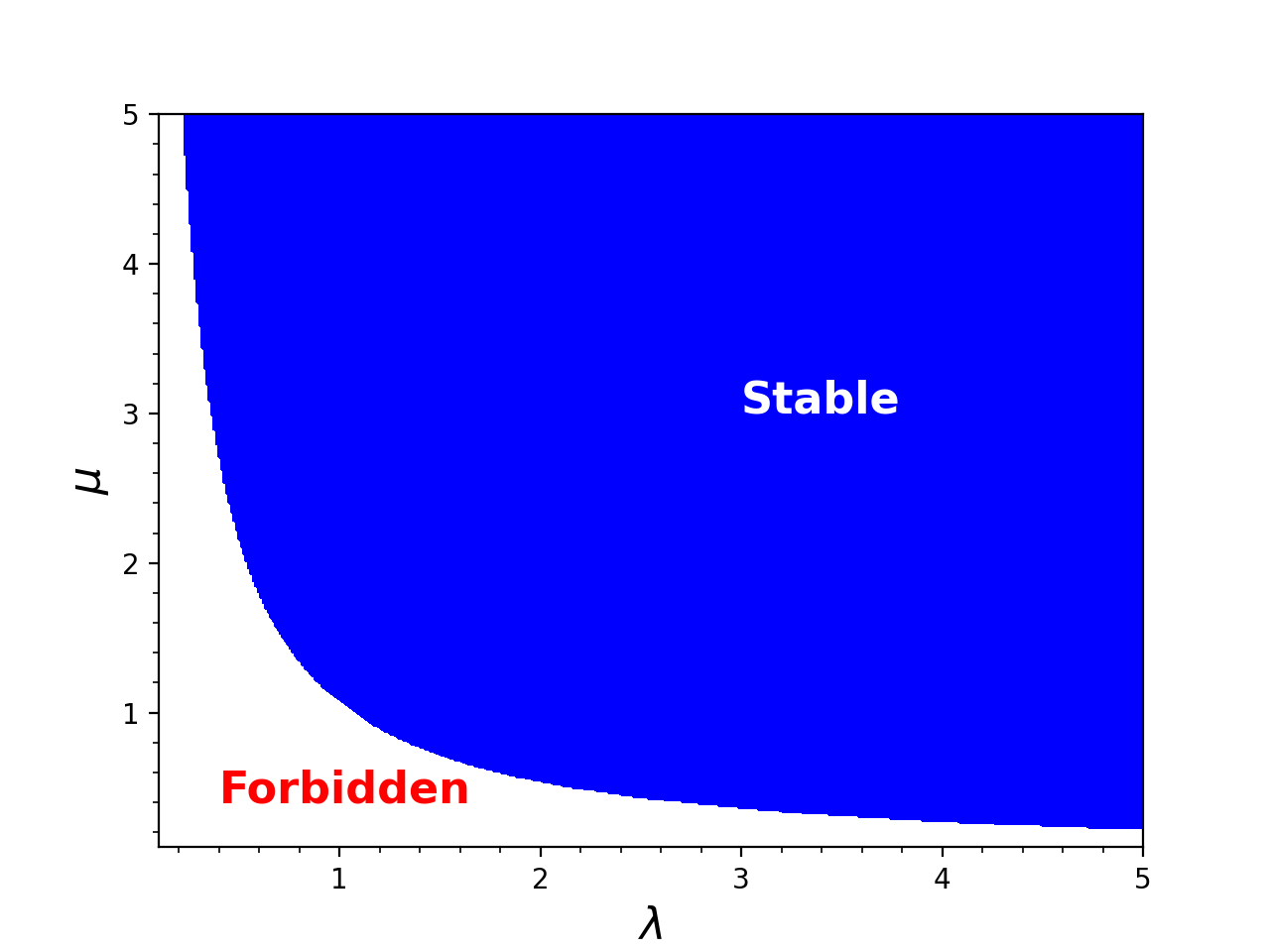}}}%
    \subfloat[\centering Density parameters]{{\includegraphics[height=5.5cm]{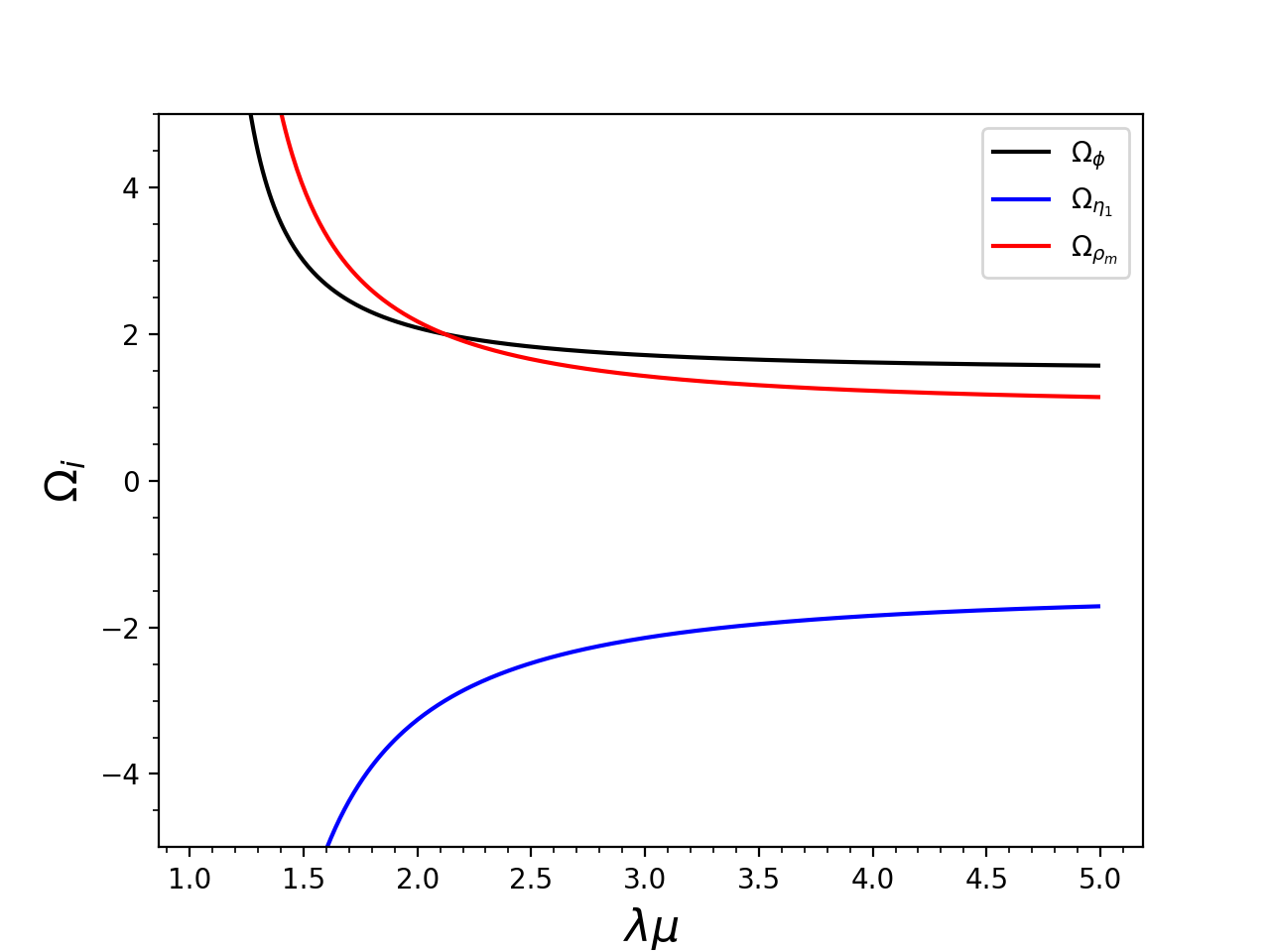}}}%
    \caption{
    On the left panel, the graph (a) shows the stability of the $\mathcal{F}_2^{(A)}$ critical point. The stable critical points lie in the blue shaded region while non-shaded region is forbidden for critical points. On the right panel, the graph (b) illustrates the relation between density parameter of non-vanishing fields and free parameters, i.e., $\lambda$ and $\mu$. The black, blue and red line in graph (b) indicates $\Omega_\phi$, $\Omega_{\eta_1}$ and $\Omega_{\rho_m}$, respectively.}
    \label{fig:plotFA2}%
\end{figure}

$\bullet$ \hypertarget{FA2}{$\mathcal{F}_2^{(A)}$}, the $X_4$ is eliminated by using Eq.(\ref{eq:contA}). The critical point is determined as
\begin{align}
    X_2= 0,\quad X_5= \lambda,\quad X_7= \frac{2 \mu }{\sqrt{\frac{8 \lambda ^2 \mu ^2}{3}-3}},\quad
    X_9= \frac{\sqrt{8 \lambda ^2 \mu ^2+18}}{\sqrt{\lambda  \left(8 \lambda ^2 \mu ^2-9\right)}}.
\end{align}
Having use Eq.(\ref{eq:eigenv}), the eigenvalues of the critical point are obtained. As a result, the eigenvalues are very complicated form and cannot be written in the analytical form. We therefore study them numerically. The stability of this critical point behaves as a stable node for all positive $\lambda$ and $\lambda\mu > \frac{3}{2\sqrt{2}}$. The plot of the stable region in $\lambda-\mu$ plane is depicted in Fig.(\ref{fig:plotFA2}). While the effective equation of state of this system then simply reduces to
\begin{equation}
    w_\text{eff}=0.
\end{equation}
Therefore, the {$\mathcal{F}_2^{(A)}$} system behaves as late time dominated DM in the universe. The results of the density parameters are shown below,
\begin{align}
\Omega _{\phi }=\frac{12 \lambda ^2 \mu ^2}{8 \lambda ^2 \mu ^2-9},\quad 
\Omega _{\eta _1}=\frac{12 \lambda ^2 \mu ^2+27}{9-8 \lambda ^2 \mu ^2},\quad
\Omega _{\rho _m}=\frac{2 \left(4 \lambda ^2 \mu ^2+9\right)}{8 \lambda ^2 \mu ^2-9}.
\end{align}
\newpage
\subsubsection{$\mathcal{F}_i^{(B)}$ : $(\phi,\eta_1,A,\rho_m)$-system (B)}
In this sector, we consider $(\phi,\eta_1,A,\rho_m)$-system case i.e., $X_3=X_8=0$. The constraint equation becomes 
\begin{equation}
1= X_2^2 - X_4^2 - \frac{X_1^2}{2 X_5^2} + X_5^2 X_7^2 - X_6^2 X_7^2 + X_5 X_9^2 - \lambda^2X_7^2. \label{eq:contB}
\end{equation}
The critical points of this system are listed below.\\

\begin{figure}[H]%
    \centering
    \subfloat[\centering Stability]{{\includegraphics[height=5.5cm]{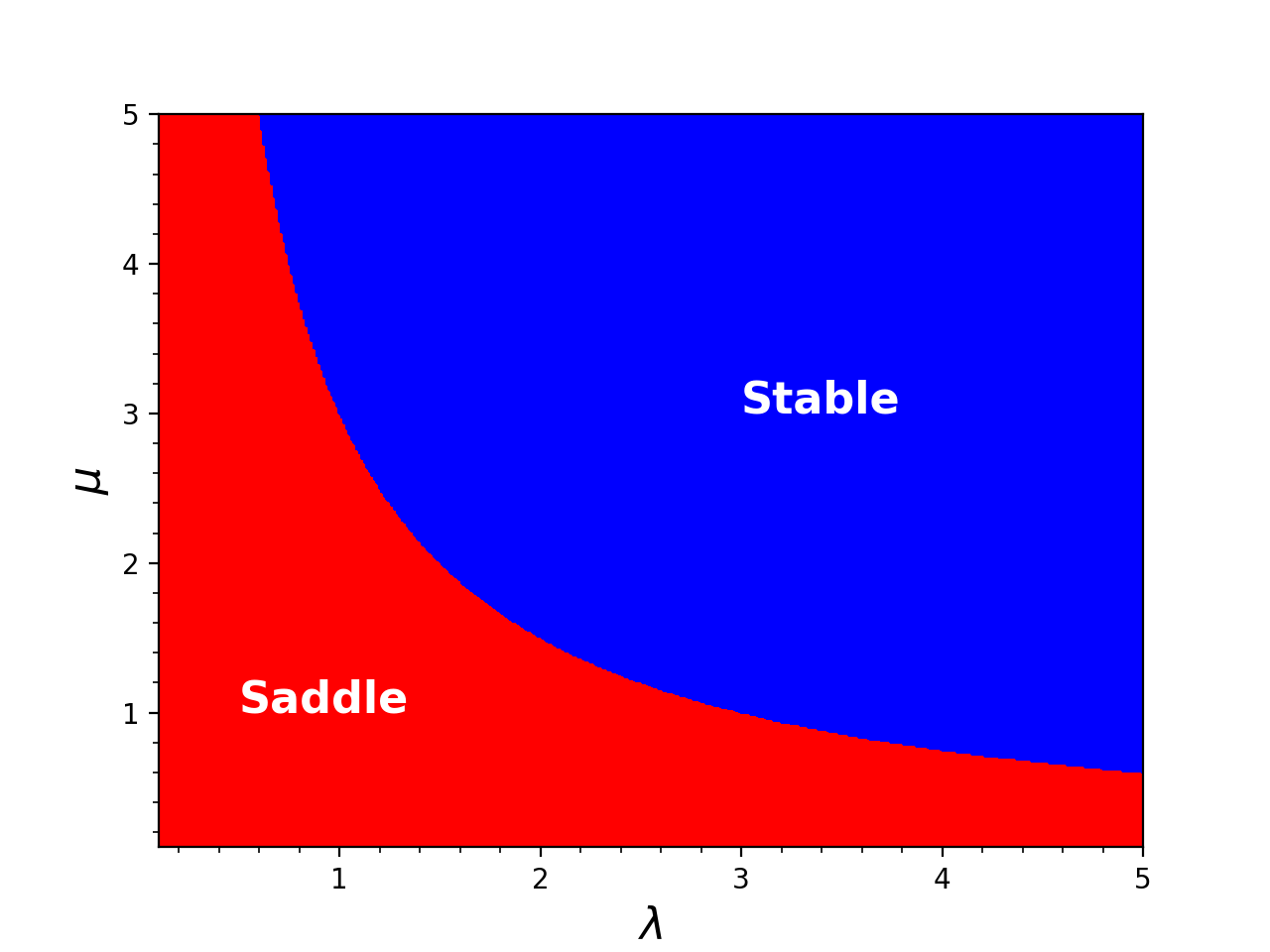}}}%
    \subfloat[\centering Effective equation of state]{{\includegraphics[height=5.5cm]{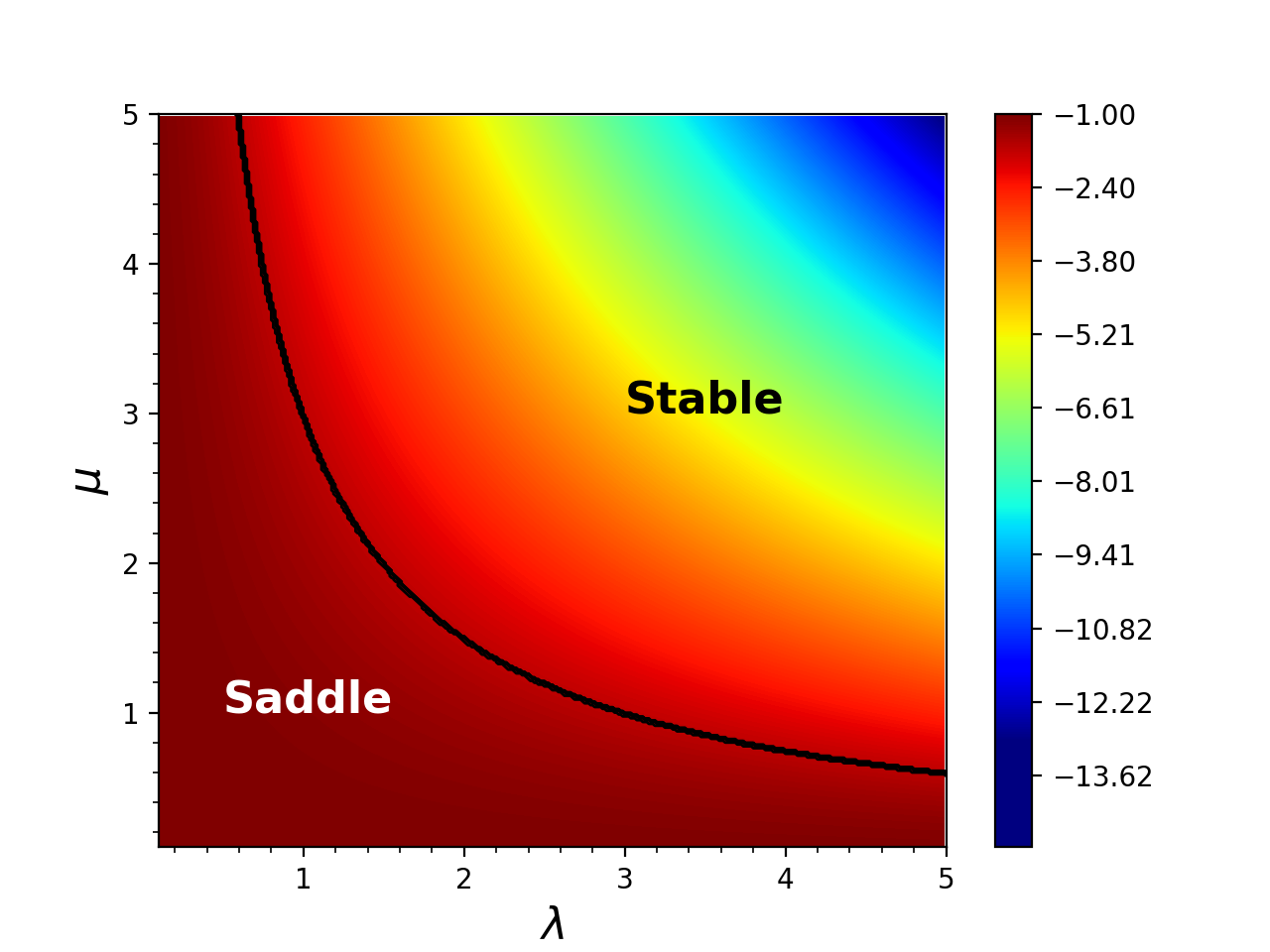}}}%
    \caption{
    On the left panel, the graph (a) shows the stability of the $\mathcal{F}_2^{(B)}$ critical point. The saddle critical points lie in the red shaded region while blue shaded region, $\lambda > 3/\mu$, is the region where critical points are stable. On the right panel, the graph (b) illustrates the changes of effective equation of state over $\lambda$ and $\mu$. Noted that the black line in graph (b) indicates the border between saddle and stable critical points.}
    \label{fig:plotFB2}%
\end{figure}

$\bullet$ \hypertarget{FB1}{$\mathcal{F}_1^{(B)}$}, we get rid of the $X_6$ variable by using constraint equation Eq.(\ref{eq:contB}). By doing this, the system provides 2 solutions of critical point. The first solution is given by 
\begin{align}
    X_1=&0,\quad X_2=0,\quad X_4=-\frac{3 \sqrt{3}}{\sqrt{8 \lambda ^2 \mu ^2+9}},\quad X_5=\frac{\sqrt{8 \lambda ^2 \mu ^2+9}}{2 \sqrt{2} \mu }\nonumber \\
     X_7=&\frac{2 \sqrt{3} \mu }{\sqrt{8 \lambda ^2 \mu ^2+9}},\quad X_9=2^{3/4} \sqrt{\frac{\mu  \left(8 \lambda ^2 \mu ^2+63\right)}{\left(8 \lambda ^2 \mu ^2+9\right)^{3/2}}}.
\end{align}
We note that the existence of the critical point exists for all positive $\lambda$ and $\mu$. The eigenvalues of this critical points cannot write in the close form and numerical study is required. The stability analysis shows that this critical point is always saddle point. The effective equation of state of this system is given by
\begin{equation}
    w_\text{eff}=0.
\end{equation}
According to the dynamical system analysis, $\mathcal{F}_1^{(B)}$ can behave like DM phase or matter dominated era of the universe. Finally, the density parameters of the system are read,
\begin{align}
\Omega _A=&-\frac{81}{16 \lambda ^2 \mu ^2+18},\quad \Omega _{\phi }=\frac{3}{2},\nonumber\\
\Omega _{\eta _1}=&-\frac{3 \left(4 \lambda ^2 \mu ^2+9\right)}{8 \lambda ^2 \mu ^2+9}, \quad \Omega _{\rho _m}=\frac{8 \lambda ^2 \mu ^2+63}{8 \lambda ^2 \mu ^2+9}.
\end{align}
\newpage
\begin{figure}[H]%
    \centering
    \includegraphics[height=5.5cm]{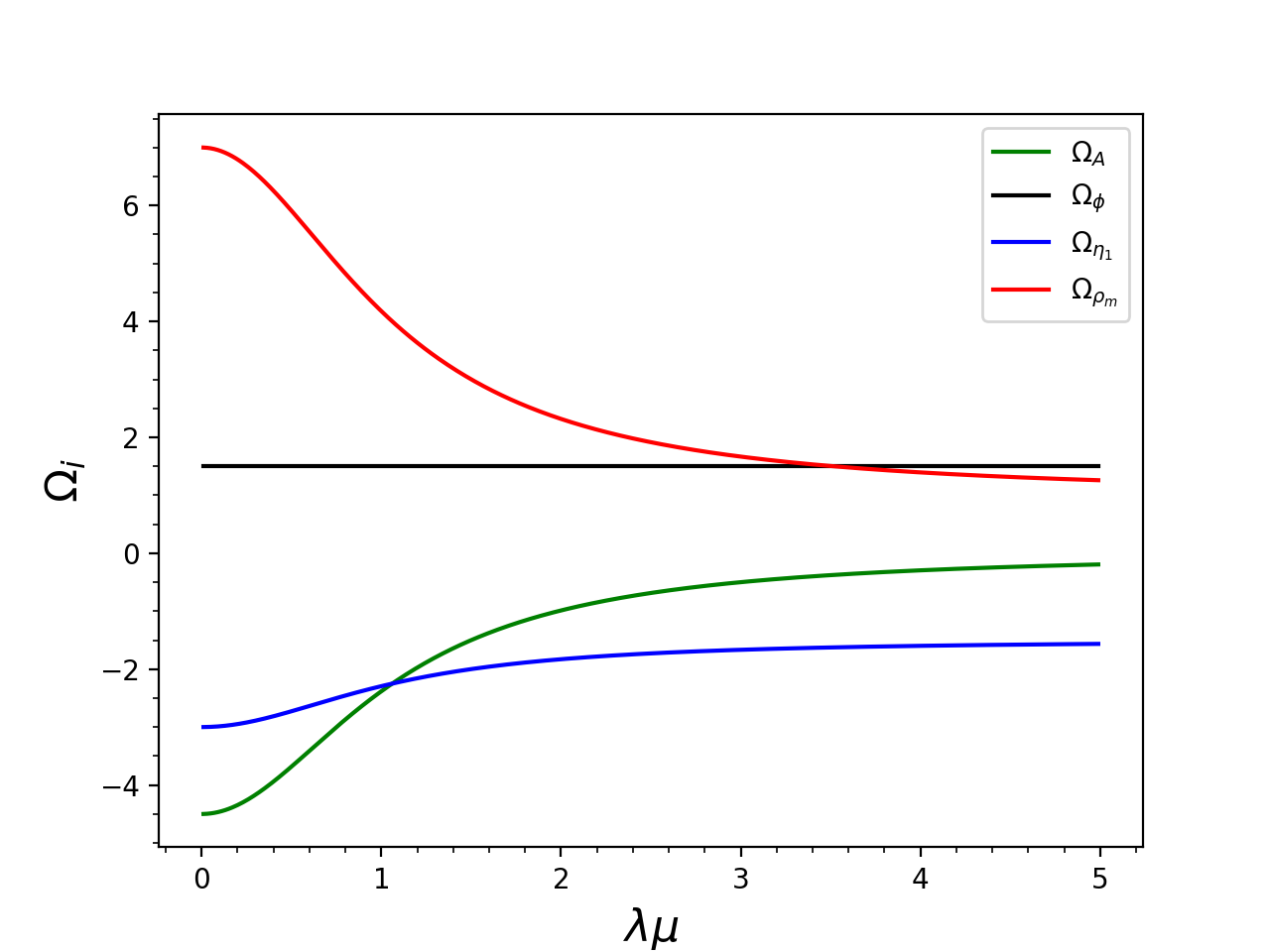}%
    \caption{
    The figure illustrates the relation between density parameter of non-vanishing fields and free parameters, i.e., $\lambda$ and $\mu$, for $\mathcal{F}_1^{(B)}$ critical point. Noted that the green, black, blue and red line in the graph indicates $\Omega_A$, $\Omega_\phi$, $\Omega_{\eta_1}$ and $\Omega_{\rho_m}$, respectively.
    }
    \label{fig:plotFB1}%
\end{figure}
$\bullet$ \hypertarget{FB2}{$\mathcal{F}_2^{(B)}$}, this is the second solution of the $(\phi,\eta_1,A,\rho_m)$ system where we eliminate $X_6$. It is written by
\begin{align}
    X_1=&0,\quad X_2=0,\quad X_9=0,\nonumber \\
    X_4=&\frac{1}{12 \lambda  \mu }\left(5 \sqrt{3}-\sqrt{8 \lambda ^2 \mu ^2+75}\right) \sqrt{\sqrt{3} \sqrt{8 \lambda ^2 \mu ^2+75}+15},\nonumber \\
    X_5=&-\frac{1}{2 \sqrt{3}}\sqrt{\frac{80 \lambda ^2 \mu ^2-61 \sqrt{3} \sqrt{8 \lambda ^2 \mu ^2+75}+915}{\mu ^2}},\nonumber \\
    X_7=&-\frac{1}{2 \lambda }\sqrt{\sqrt{\frac{8 \lambda ^2 \mu ^2}{3}+25}+5}.
\end{align}
The eigenvalues of the $\mathcal{F}_2^{(B)}$ system are very complicated. By numerical study, the stability of this critical point can be both stable and saddle points for all positive $\lambda$ and $\mu$. In addition, their existences of the stable node in the $\lambda-\mu$ plane are shown in Fig.(\ref{fig:plotFB2}). The effective equation of state of this system can be written as
\begin{equation}
    w_\text{eff}=\frac{1}{9} \left(6-\sqrt{3} \sqrt{8 \lambda ^2 \mu ^2+75}\right).
\end{equation}
The density parameters of the system are shown below
\begin{align}
\Omega_A=&\frac{1}{12} \left(39-5 \sqrt{3} \sqrt{8 \lambda ^2 \mu ^2+75}\right),\nonumber\\
\Omega_{\phi }=&\frac{5}{3} \sqrt{\frac{8 \lambda ^2 \mu ^2}{3}+25}-\frac{11}{6},\nonumber\\
\Omega_{\eta _1}=&-\frac{5}{36} \left(\sqrt{3} \sqrt{8 \lambda ^2 \mu ^2+75}+3\right),\nonumber\\
\Omega_{\rho _m}=&0.
\end{align}

\begin{figure}[H]%
    \centering
    \includegraphics[height=5.5cm]{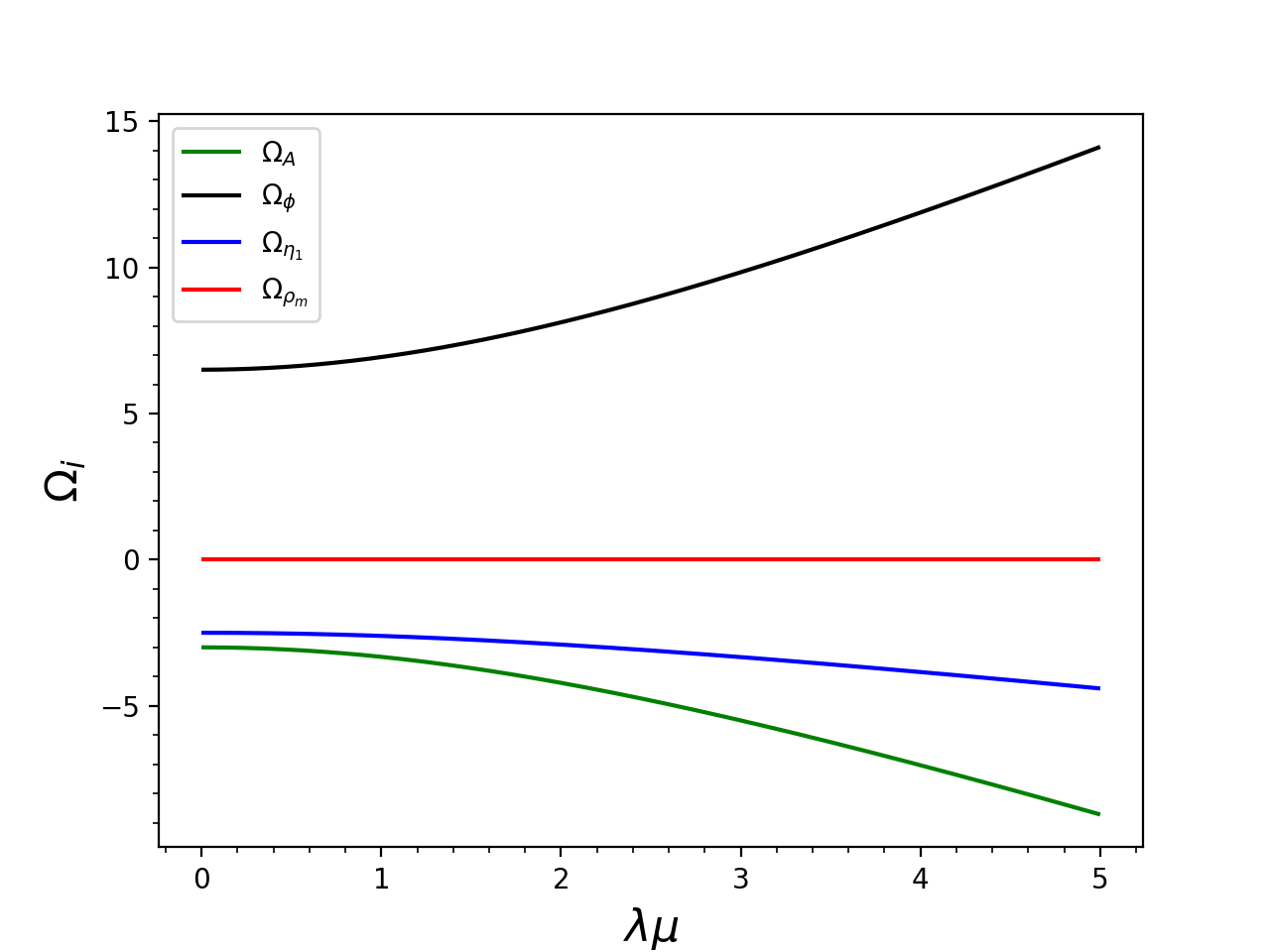}%
    \caption{
    The figure illustrates the relation between density parameter of non-vanishing fields and free parameters, i.e., $\lambda$ and $\mu$, for $\mathcal{F}_2^{(B)}$ critical point. Noted that the green, black, blue and red line in the graph indicates $\Omega_A$, $\Omega_\phi$, $\Omega_{\eta_1}$ and $\Omega_{\rho_m}$, respectively.
    }
    \label{fig:plotFB2_2}%
\end{figure}

\begin{figure}[H]%
    \centering
    \subfloat[\centering Stability]{{\includegraphics[height=5.5cm]{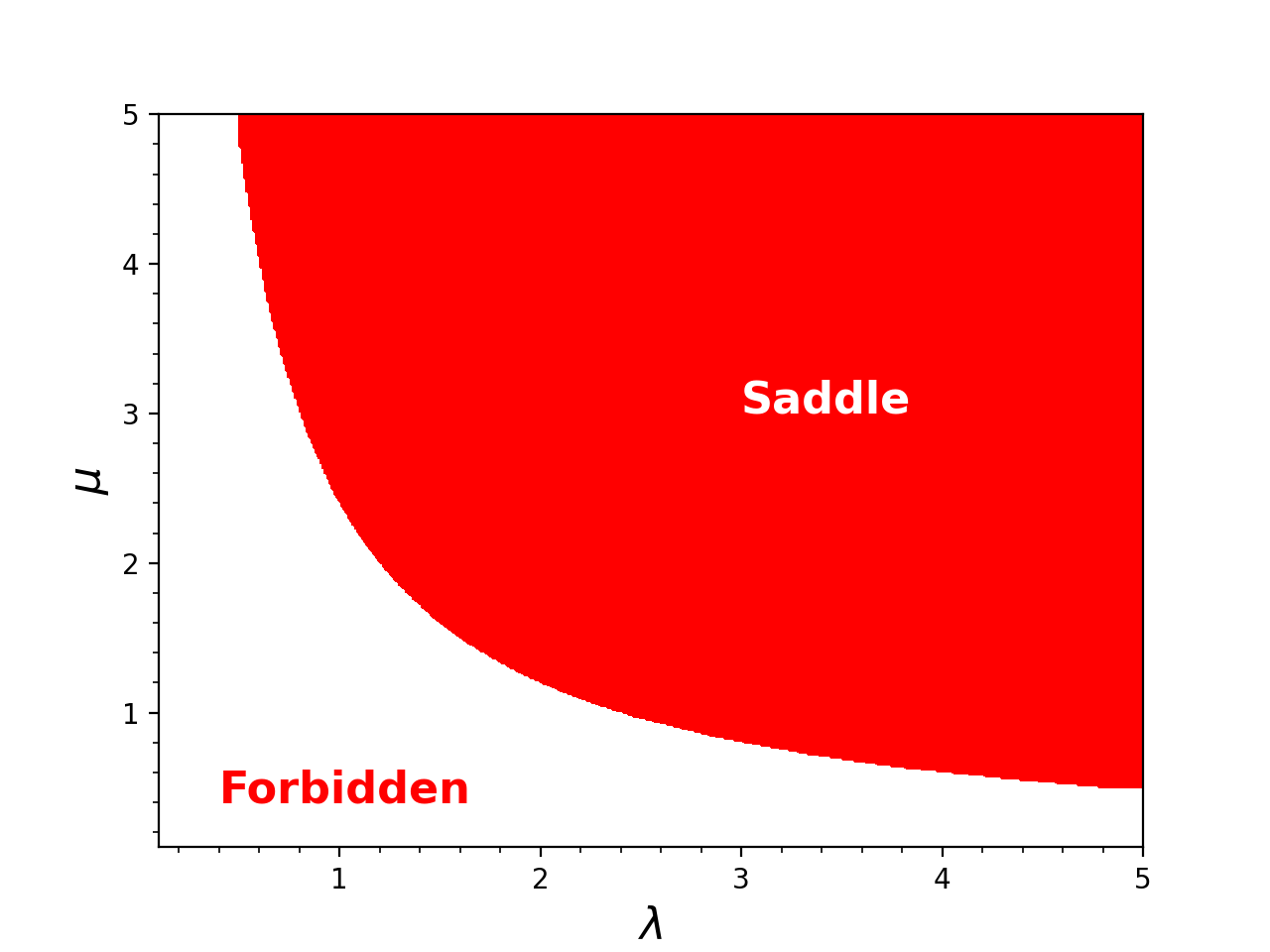}}}%
    \subfloat[\centering Density parameters]{{\includegraphics[height=5.5cm]{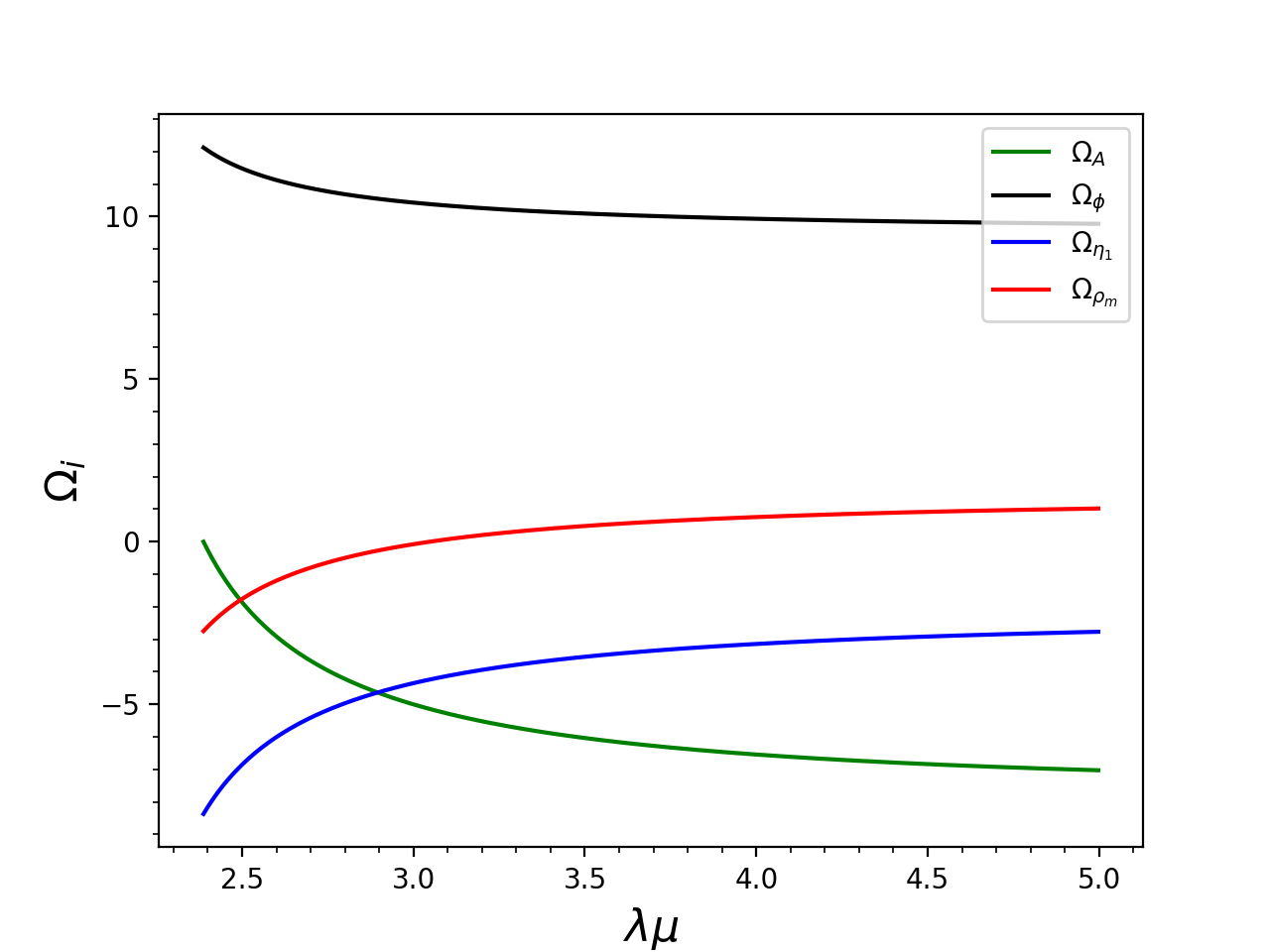}}}%
    \caption{
    On the left panel, the graph (a) shows the stability of the $\mathcal{F}_3^{(B)}$ critical point. The saddle critical points lie in the red shaded region while non-shaded region is forbidden for critical points. On the right panel, the graph (b) illustrates the relation between density parameter of non-vanishing fields and free parameters, i.e., $\lambda$ and $\mu$. The green, black, blue and red line in graph (b) indicates $\Omega_A$, $\Omega_\phi$, $\Omega_{\eta_1}$ and $\Omega_{\rho_m}$, respectively.}
    \label{fig:plotFB3}%
\end{figure}

$\bullet$ \hypertarget{FB3}{$\mathcal{F}_3^{(B)}$}, the $X_9$ is excluded from the system by using Eq.(\ref{eq:contB}). We obtain the critical point as
    \begin{align}
    X_2=&0,\quad X_1=\frac{\sqrt{46 \lambda ^2 \mu ^2-\Delta-429}}{2 \sqrt{10} \mu }, \quad X_4= \frac{\sqrt{\mu ^2 \left(234-56 \lambda ^2 \mu ^2 - \Delta\right)}}{\mu  \sqrt{12 \lambda ^4 \mu ^4-68 \lambda ^2 \mu ^2+63}},\nonumber \\
    X_5=& \frac{1}{2 \sqrt{10}}\sqrt{\frac{86 \lambda ^2 \mu ^2-\Delta-269}{\mu ^2}},\quad X_6=\frac{\sqrt{46 \lambda ^2 \mu ^2-\Delta-429}}{2 \sqrt{10} \mu },\nonumber \\
    X_7=&\frac{\sqrt{\mu ^2 \left(234-56 \lambda ^2 \mu ^2-\Delta\right)}}{\sqrt{12 \lambda ^4 \mu ^4-68 \lambda ^2 \mu ^2+63}},
\end{align}
where
\begin{align}
    \Delta=\sqrt{6916 \lambda ^4 \mu ^4-47628 \lambda ^2 \mu ^2+74601}. \label{eq:Delta}
\end{align}
The eigenvalues of this system are also very complicated. With help from the numerical analysis, this critical point is always saddle point and the existence of the saddle point exists in the range of positive $\lambda$ and $\lambda\mu>\sqrt{\frac{57}{10}}$. The plot is depicted in Fig.(\ref{fig:plotFB3}). The effective equation of state of this system is given by
\begin{equation}
    w_\text{eff}=-\frac{5}{3}.
\end{equation}
The non-vanishing energy density parameters are read
\begin{align}
\Omega _A=&\frac{\left(429-46 \lambda ^2 \mu ^2-\Delta\right)\left(78 \lambda ^2 \mu ^2+\Delta-369\right)}{8 \left(2 \lambda ^2 \mu ^2-9\right) \left(86 \lambda ^2 \mu ^2+\Delta-269\right)},\nonumber\\
\Omega _{\phi }=&\frac{\left(-56 \lambda ^2 \mu ^2+\Delta+234\right)\left(86 \lambda ^2 \mu ^2+\Delta-269\right)}{40 \left(12 \lambda ^4 \mu ^4-68 \lambda ^2 \mu ^2+63\right)},\nonumber\\
\Omega _{\eta _1}=&\frac{\left(\lambda ^2 \mu ^2+1\right) \left(56 \lambda ^2 \mu ^2-\Delta-234\right)}{12 \lambda ^4 \mu ^4-68 \lambda ^2 \mu ^2+63},\nonumber\\
\Omega _{\rho _m}=&\frac{436 \lambda ^4 \mu ^4+2 \lambda ^2 \mu ^2 \left(3 \Delta-2998\right)-57 \Delta+18693}{2 \left(2 \lambda ^2 \mu ^2-9\right) \left(86 \lambda ^2 \mu ^2+\Delta-269\right)}.
\end{align}
\newpage

\subsubsection{$\mathcal{F}_i^{(C)}$ : $(\phi,\eta_0,\eta_1,\rho_m)$-system (C)}
In this system, we have assumed that $X_1=0$ and $X_6=0$. The constraint equation is written by
\begin{equation}
    1 = X_2^2 - X_3^2 - X_4^2 + X_5^2 X_7^2 + X_5 X_9^2 - \lambda^2X_7^2  - \lambda^2X_8^2.  \label{eq:contC}
\end{equation}

$\bullet$ \hypertarget{FC1}{$\mathcal{F}_1^{(C)}$}, the critical point of the system is given by
\begin{align}
    X_3&= 0,\quad X_8=0, \quad X_4=\frac{3}{\sqrt{\frac{1}{3} \left(8 \lambda ^2 \mu ^2-6\right)}}, \quad X_5 =\frac{\sqrt{4 \lambda ^2 \mu ^2-9}}{2 \mu }, \nonumber\\
    X_7 &=-\frac{2 \mu }{\sqrt{\frac{1}{3} \left(8 \lambda ^2 \mu ^2-6\right)}}, \quad X_9 =\frac{2 \sqrt{\frac{27}{8 \lambda ^2 \mu ^2-6}+\frac{1}{2}}}{\sqrt{\frac{\sqrt{4 \lambda ^2 \mu ^2-9}}{\mu }}}.
\end{align}
Using definition of stability matrix in Eq.(\ref{eq:eigenv}), we obtain the eigenvalues and they are simply reduced to
\begin{equation}
    \left(0,0,0,-\frac{i}{2}  \sqrt{4 \lambda ^2 \mu ^2-9},\frac{i}{2} \sqrt{4 \lambda ^2 \mu ^2-9},-\frac{9}{5}\right)\quad \stackrel{{\rm real~ part}}{\longrightarrow} \quad\left(0,0,0,0,0,-\frac{9}{5}\right),
\end{equation}
for positive $\lambda$ and $\lambda\mu >\frac{3}{2}$. The effective equation of state reduces to zero,
\begin{equation}
    w_\text{eff}=0.
\end{equation}
This $(\phi,\eta_0,\eta_1,\rho_m)$ system behaves as DM. The density parameters are
\begin{align}
    \Omega_\phi &=\frac{3 \left(4 \lambda ^2 \mu ^2-9\right)}{8 \lambda ^2 \mu ^2-6},\quad \Omega_{\eta_0} =0,\nonumber \\
    \Omega_{\eta_1} &=\frac{3 \left(4 \lambda ^2 \mu ^2+9\right)}{6-8 \lambda ^2 \mu ^2},\quad \Omega_{\rho_m} =\frac{27}{4 \lambda ^2 \mu ^2-3}+1.
\end{align}
\begin{figure}[H]%
    \centering
    \subfloat[\centering Stability]{{\includegraphics[height=5.5cm]{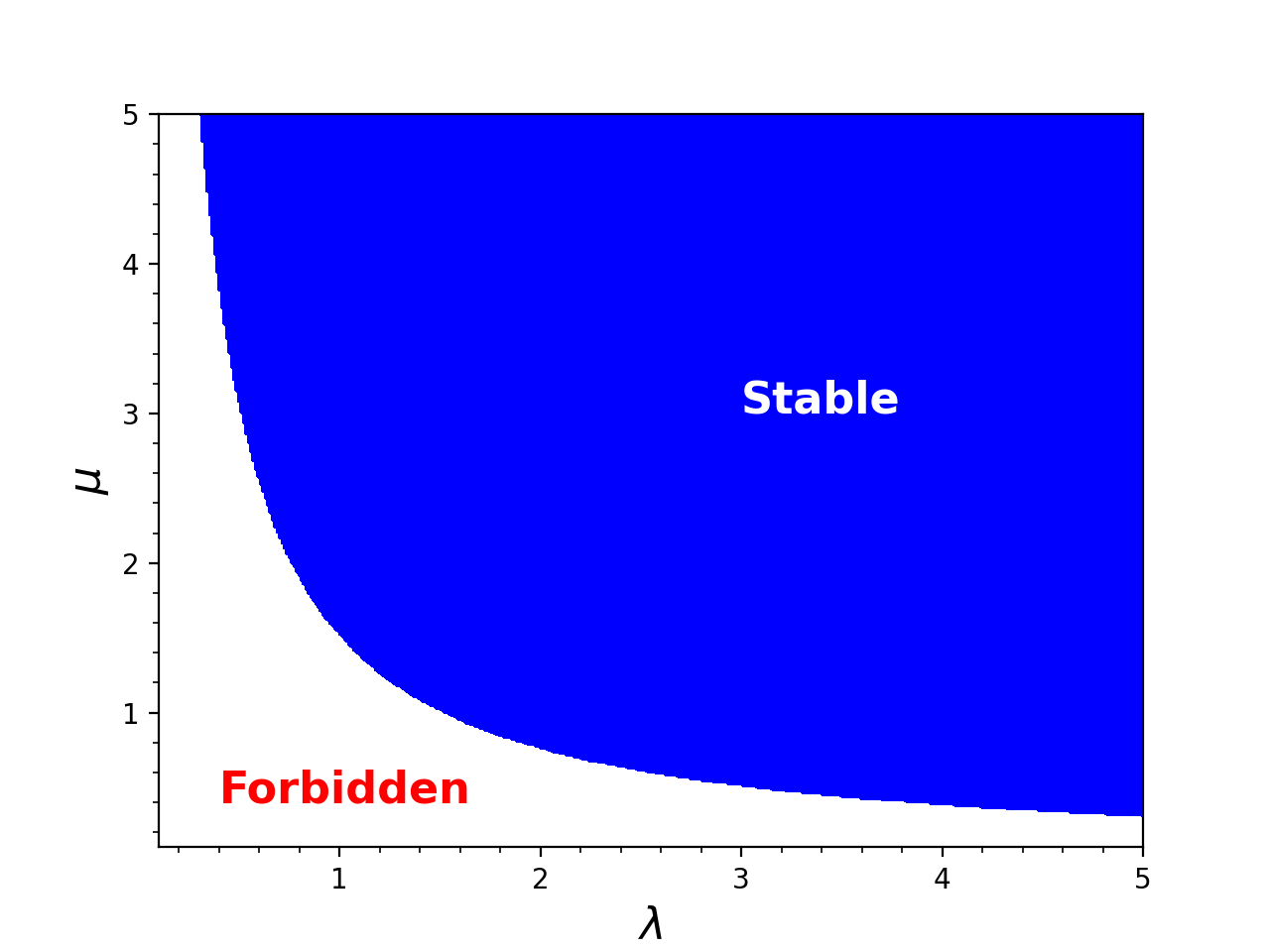}}}
    \subfloat[\centering Density parameters]{{\includegraphics[height=5.5cm]{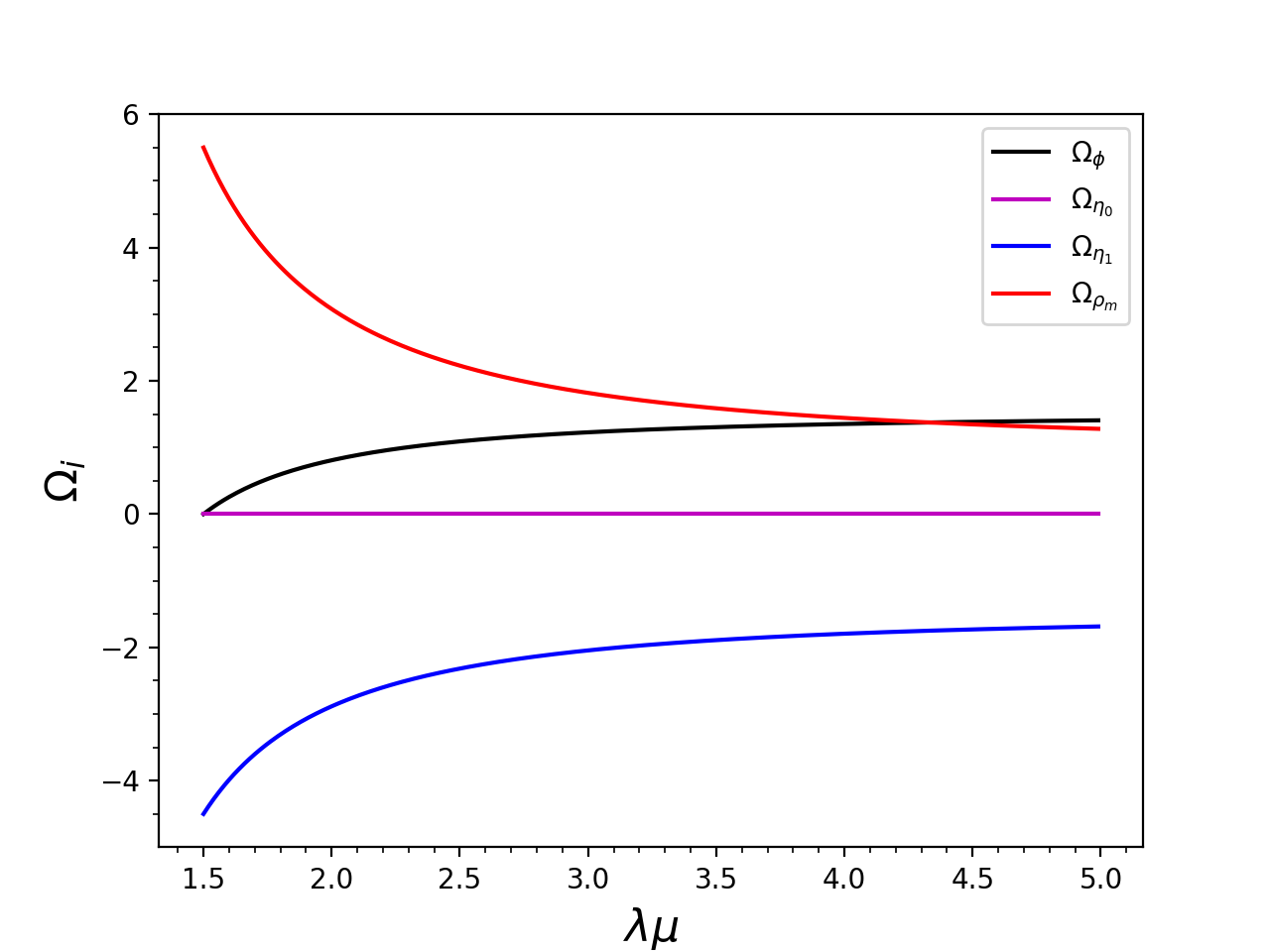}}}%
    \caption{
    On the left panel, the graph (a) shows the stability of the $\mathcal{F}_1^{(C)}$ critical point. The stable critical points lie in the blue shaded region while non-shaded region is forbidden for critical points. On the right panel, the graph (b) illustrates the relation between density parameter of non-vanishing fields and free parameters, i.e., $\lambda$ and $\mu$. The black, purple, blue and red line in graph (b) indicates $\Omega_\phi$, $\Omega_{\eta_0}$, $\Omega_{\eta_1}$ and $\Omega_{\rho_m}$, respectively.}
    
    \label{fig:plotFC1}%
\end{figure}

\subsubsection{$\mathcal{F}_i^{(D)}$ : $(\phi,\eta_0,\eta_1,A,\rho_m)$-system (D)} 
In this case, we consider that all fields in the model are not vanished. Therefore, the constraint equation reads
\begin{equation}
1 = X_2^2 - X_3^2 - X_4^2 - \frac{X_1^2}{2 X_5^2} + X_5^2 X_7^2 - X_6^2 X_7^2 + X_5 X_9^2 - \lambda^2 X7^2 - \lambda^2X_8^2. \label{eq:contD}
\end{equation}

$\bullet$ \hypertarget{FD1}{$\mathcal{F}_1^{(D)}$}, we replace the $X_6$ variable in the system by using constraint equation, Eq.(\ref{eq:contD}). The critical point is given by
\begin{align}
    X_1=&0,\quad X_2=0,\quad X_3=0,\quad X_8=0,\nonumber \\
    X_4=&\frac{3 \sqrt{3}}{\sqrt{8 \lambda ^2 \mu ^2+9}},\quad 
    X_5=\frac{\sqrt{8 \lambda ^2 \mu ^2+9}}{2 \sqrt{2} \mu },\nonumber \\
    X_7=&-\frac{2 \sqrt{3} \mu }{\sqrt{8 \lambda ^2 \mu ^2+9}},\quad
    X_9=-\frac{2^{3/4} \sqrt{\mu  \left(8 \lambda ^2 \mu ^2+63\right)}}{\left(8 \lambda ^2 \mu ^2+9\right)^{3/4}}.
\end{align}
The eigenvalues of this system are lengthy and complicated. According to the numerical study, the eigenvalues can be zero, positive and negative real numbers for all positive $\lambda$ and $\mu$. This means that the critical point is always the saddle point. While the effective equation of state of this system reads
\begin{align}
    w_\text{eff}=0.
\end{align}
As a result, the critical point of this system behaves like DM phase of matter dominated epoch of the observed universe. The density parameters are given by
\begin{align}
\Omega _A=&-\frac{81}{16 \lambda ^2 \mu ^2+18},\quad \Omega _{\phi }=\frac{3}{2},\quad \Omega _{\eta _0}=0,\nonumber\\
\Omega _{\eta _1}=&-\frac{3 \left(4 \lambda ^2 \mu ^2+9\right)}{8 \lambda ^2 \mu ^2+9},\quad \Omega _{\rho _m}=\frac{8 \lambda ^2 \mu ^2+63}{8 \lambda ^2 \mu ^2+9}.
\end{align}
\begin{figure}[H]%
    \centering
    \includegraphics[height=5.5cm]{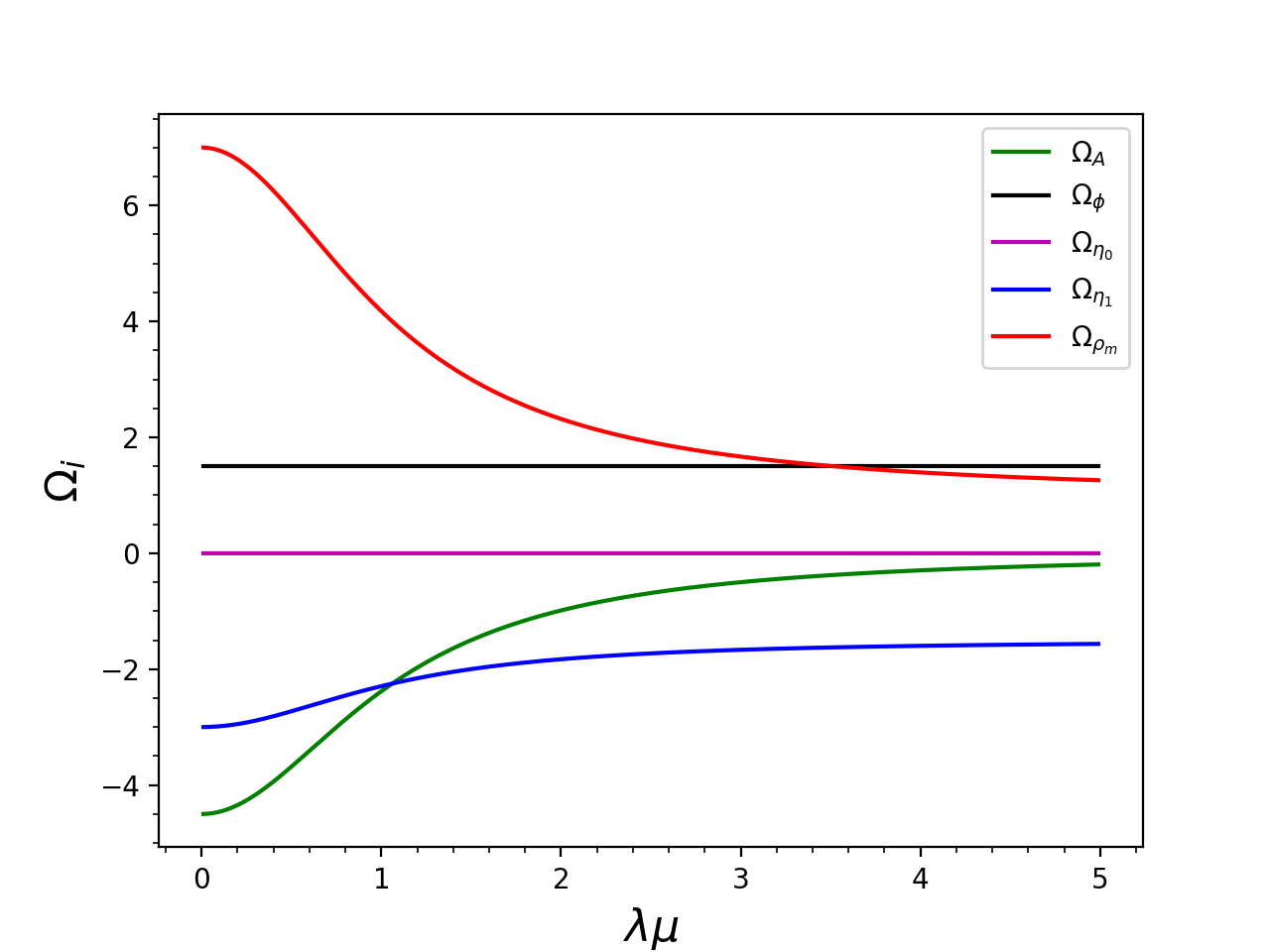}
    \caption{The figure illustrates the relation between density parameter of non-vanishing fields and free parameters, i.e., $\lambda$ and $\mu$, for $\mathcal{F}_1^{(D)}$ critical point. Noted that the green, black, purple, blue and red line in the graph indicates $\Omega_A$, $\Omega_\phi$, $\Omega_{\eta_0}$, $\Omega_{\eta_1}$ and $\Omega_{\rho_m}$, respectively.
    }
    \label{fig:plotFD1}%
\end{figure}
\newpage
$\bullet$ \hypertarget{FD2}{$\mathcal{F}_2^{(D)}$}, we eliminate the $X_9$ from the autonomous system by using Eq.(\ref{eq:contD}). The critical point is determined and it reads
\begin{align}
    X_2=&0,\quad X_3=0,\quad X_8=0,\quad X_1=\frac{\sqrt{\Delta+46 \lambda ^2 \mu ^2-429}}{2 \sqrt{10} \mu }, \nonumber \\
    X_4=&\frac{\sqrt{\mu ^2 \left(\Delta-56 \lambda ^2 \mu ^2+234\right)}}{\mu  \sqrt{12 \lambda ^4 \mu ^4-68 \lambda ^2 \mu ^2+63}},\quad X_5=\frac{1}{2 \sqrt{10}}\sqrt{\frac{\Delta+86 \lambda ^2 \mu ^2-269}{\mu ^2}},\nonumber \\ X_6=&\frac{\sqrt{\Delta+46 \lambda ^2 \mu ^2-429}}{2 \sqrt{10} \mu },\quad X_7=\frac{\sqrt{\mu ^2 \left(\Delta-56 \lambda ^2 \mu ^2+234\right)}}{\sqrt{12 \lambda ^4 \mu ^4-68 \lambda ^2 \mu ^2+63}},
\end{align}
where $\Delta$ is defined by Eq.(\ref{eq:Delta}). Due to the complexity, the eigenvalues are studied by numerical calculation. The critical point turns out to be saddle point for the range of $\lambda > 0$ and $\lambda\mu >\sqrt{\frac{57}{10}}$. The effective equation of state of this system is given by
\begin{equation}
    w_\text{eff}=-\frac{5}{3}.
\end{equation}
In addition, the density parameters are written in the following forms,
\begin{align}
\Omega _A=&\frac{\left(\Delta+46 \lambda ^2 \mu ^2-429\right) \left(-5 \Delta \left(6 \lambda ^2 \mu ^2-7\right)-\Delta^2+4576 \lambda ^4 \mu ^4-33828 \lambda ^2 \mu ^2+61686\right)}{40 \left(12 \lambda ^4 \mu ^4-68 \lambda ^2 \mu ^2+63\right) \left(\Delta+86 \lambda ^2 \mu ^2-269\right)},\nonumber\\
\Omega _{\phi}=&\frac{\left(\Delta-56 \lambda ^2 \mu ^2+234\right) \left(\Delta+86 \lambda ^2 \mu ^2-269\right)}{40 \left(12 \lambda ^4 \mu ^4-68 \lambda ^2 \mu ^2+63\right)},\nonumber\\
\Omega _{\eta _0}=&0,\nonumber \\
\Omega _{\eta_1}=&\frac{\left(\lambda ^2 \mu ^2+1\right) \left(-\Delta+56 \lambda ^2 \mu ^2-234\right)}{12 \lambda ^4 \mu ^4-68 \lambda ^2 \mu ^2+63},\nonumber\\
\Omega _{\rho_m}=&\frac{\Delta \left(36 \lambda ^4 \mu ^4-384 \lambda ^2 \mu ^2+399\right)-6 \Delta^2+2616 \lambda ^6 \mu ^6+2468 \lambda ^4 \mu ^4-131638 \lambda ^2 \mu ^2+316755}{2 \left(12 \lambda ^4 \mu ^4-68 \lambda ^2 \mu ^2+63\right) \left(\Delta+86 \lambda ^2 \mu ^2-269\right)}.
\end{align}
\newpage
\begin{figure}[H]%
    \centering
    \subfloat[\centering Stability]{{\includegraphics[height=5.5cm]{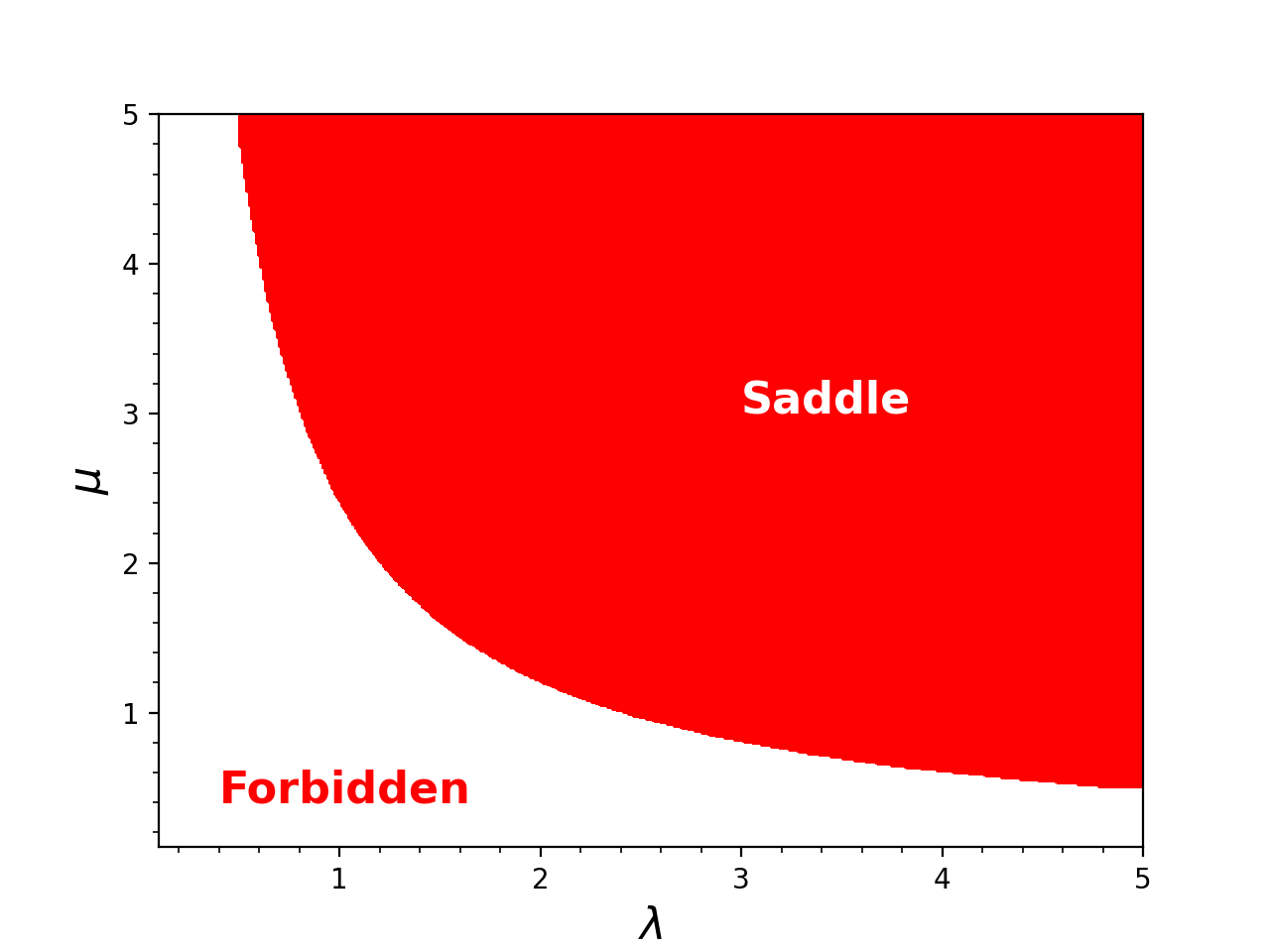}}}%
    \subfloat[\centering Density parameters]{{\includegraphics[height=5.5cm]{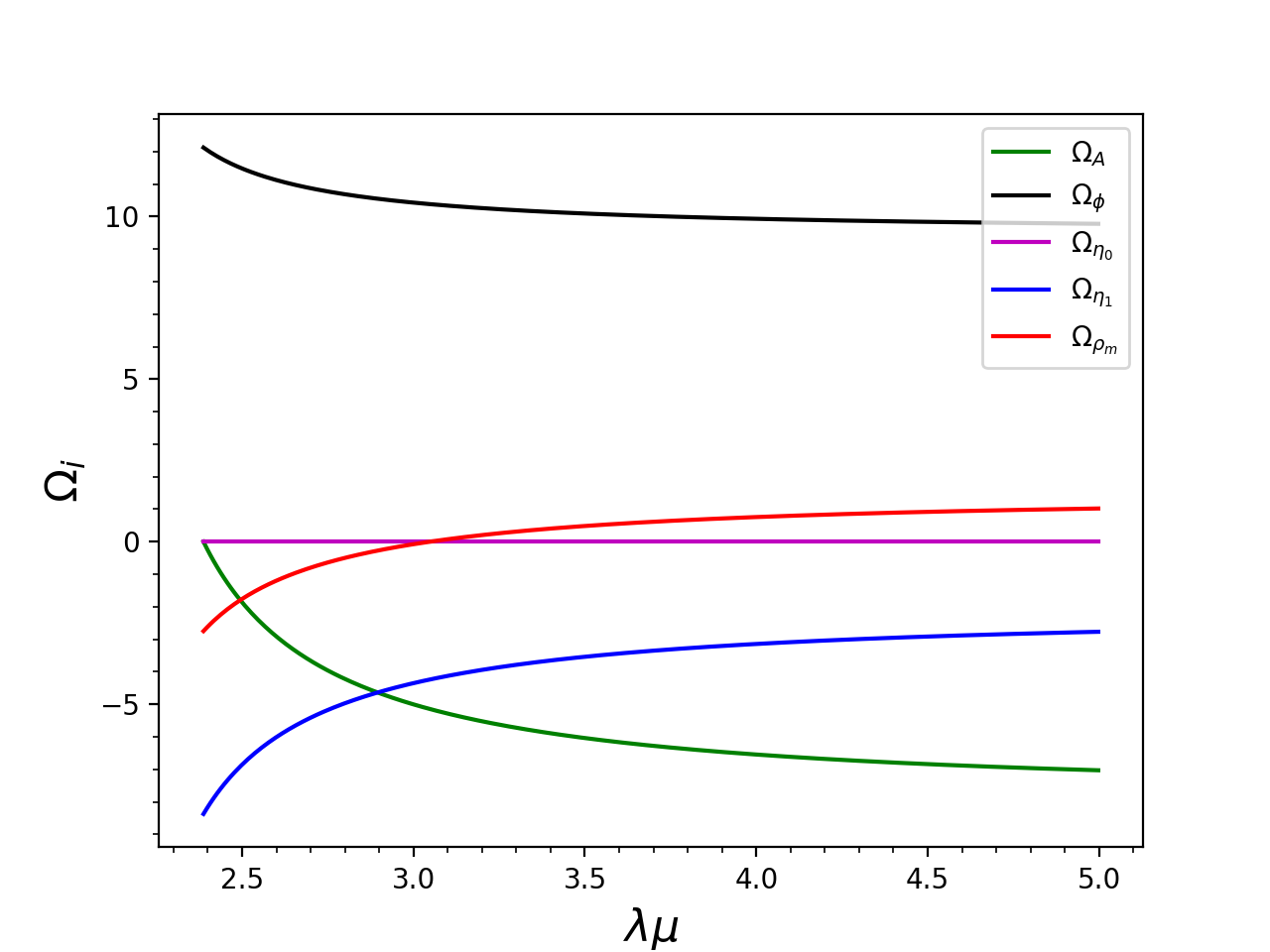}}}%
    \caption{On the left panel, the graph (a) shows the stability of the $\mathcal{F}_2^{(D)}$ critical point. The saddle critical points lie in the red shaded region while non-shaded region is forbidden for critical points. On the right panel, the graph (b) illustrates the relation between density parameter of non-vanishing fields and free parameters, i.e., $\lambda$ and $\mu$. The green, black, purple, blue and red line in graph (b) indicates $\Omega_A$, $\Omega_\phi$, $\Omega_{\eta_0}$, $\Omega_{\eta_1}$ and $\Omega_{\rho_m}$, respectively.}
    
    \label{fig:plotFD2}%
\end{figure}

\subsection{\label{sec:MB5!=0}$M_{(5)}\neq0$ case with slow-roll scalar fields}
From now on, we will assume that the slow-roll approximation is used in order to evaluate the critical point in this case. In the other word, the kinetic term of $\eta_0$ and $\eta_1$ are vanished ($X_4=0$ and $X_3=0$), if any of these fields exists in the considering system. Contrary to the previous subsection, the dilaton field can be decoupled from the autonomous.
\subsubsection{$\bar{\mathcal{F}}_i^{(B)}$ : $(\phi,\eta_1,A,\rho_m)$-system (B)}
In this case, we consider $(\phi,\eta_1,A,\rho_m)$-system and this leads to $X_3=0$, $X_4=0$ and $X_8=0$. The constraint equation is given by
\begin{equation}
    1 = X_2^2 - \frac{X_1^2}{2 X_5^2} + X_5^2 X_7^2 - X_6^2 X_7^2 + X_5 X_9^2 - \lambda^2X_7^2 \label{eq:contBs}
\end{equation}

$\bullet$ \hypertarget{FBbar1}{$\bar{\mathcal{F}}_1^{(B)}$}, the $X_1$ variable is eliminated by using Eq.(\ref{eq:contBs}). The critical point is read
\begin{align}
X_2=&0,\quad X_9=0,\nonumber \\
    X_5=& \sqrt{\frac{1}{458} \left(29 \sqrt{273}+547\right)} \lambda,\nonumber \\
    X_6 =& \sqrt{\frac{2}{229} \left(7 \sqrt{273}-18\right)} \lambda ,\nonumber \\
    X_7 =& \sqrt{\frac{3 \sqrt{273}}{8 \lambda ^2}-\frac{25}{8 \lambda ^2}}.
\end{align}
The eigenvalues have been studied with numerical calculation due to their complication. The result shows that the critical point is always saddle point for all positive $\lambda$ and $\mu$. Effective equation of state is written by
\begin{equation}
    w_\text{eff} =-1.
\end{equation}
As a result, we might interpret that the critical point can be identified as DE (de-Sitter) phase at the late time. Non-vanishing density parameters are given by
\begin{align}
    \Omega_A &=\frac{1}{2} \left(\sqrt{273}-21\right),\quad \Omega_\phi =\frac{1}{4} \left(\sqrt{273}+11\right),\nonumber \\
    \Omega_{\eta_1} &=\frac{1}{8} \left(25-3 \sqrt{273}\right), \quad \Omega_{\rho_m} =0.    
\end{align}

$\bullet$ \hypertarget{FBbar2}{$\bar{\mathcal{F}}_2^{(B)}$}, we use constraint equation in Eq.(\ref{eq:contBs}) to get rid of $X_2$ from the system. Then, the critical point of this system is
\begin{align}
    X_1 =0,\quad X_6=0,\quad X_9=0,\quad
    X_5 =\frac{\sqrt{31}\lambda}{5},\quad X_7=\frac{5}{\sqrt{6}\lambda}.
\end{align}
The eigenvalues are given by
\begin{equation}
    \left(0,-\frac{24}{5},-\frac{3}{2},-1,1\right),
\end{equation}
According to the eigenvalues given above, it clearly shows that the critical point is the saddle point. The effective equation of state of the system takes the following form, 
\begin{equation}
w_\text{eff}=-1.
\end{equation}
The critical point can behave as DE but the universe is not dominated by DE at the late time. The density parameters are written by
\begin{align}
    \Omega_A =0,\quad \Omega_\phi =\frac{31}{6},\quad
    \Omega_{\eta_1} =-\frac{25}{6}, \quad \Omega_{\rho_m} =0.
\end{align}

$\bullet$ \hypertarget{FBbar3}{$\bar{\mathcal{F}}_3^{(B)}$}, we exclude the $X_6$ from the system via the constraint equation in (\ref{eq:contBs}). The critical point is determined by the following form, 
\begin{align}
    X_1=0,\quad X_2=0,\quad X_9=0,\quad
    X_5 =\sqrt{\frac{13}{5}} \lambda,\quad X_7 =\frac{1}{\lambda }\sqrt\frac{5}{2}.
\end{align}
The real part of eigenvalues is read
\begin{eqnarray}
    &&\left(-\frac{3}{2},1,-4.59,-1.60-3.56 i,-1.60+3.56 i\right)\nonumber\\\quad \stackrel{{\rm real~part}}{\longrightarrow}\quad &&\left(-\frac{3}{2},1,-4.59,-1.60,-1.60\right).
\end{eqnarray}
The results show that this critical point always represents saddle point for all positive $\lambda$ and $\mu$. The effective equation of state is reduced to 
\begin{equation}
w_\text{eff}=-1.
\end{equation}
The density parameters can be written as
\begin{align}
    \Omega_A =-3,\quad \Omega_\phi =\frac{13}{2},\quad
    \Omega_{\eta_1} =-\frac{5}{2},\quad \Omega_{\rho_m} =0.
\end{align}

$\bullet$ \hypertarget{FBbar4}{$\bar{\mathcal{F}}_4^{(B)}$}, we find two solutions via eliminating the $X_7$ by Eq.(\ref{eq:contBs}). The first solution reads
\begin{align}
    X_1 =0,\quad X_2=0,\quad X_6=0,\quad X_9=0,\quad X_5 =\sqrt{\frac{7}{5}} \lambda.
\end{align}
The eigenvalues of stability matrix of this critical points are given by
\begin{equation}
    \left(0,-\frac{24}{5},-\frac{3}{2},-1,1\right).
\end{equation}
As a result, this critical point is always saddle point for all positive $\lambda$ and $\mu$. The effective equation of state is determined as 
\begin{equation}
w_\text{eff}=-\frac{1}{3}.
\end{equation}
Non-vanishing density parameters are written by
\begin{align}
    \Omega_A =0,\quad \Omega_\phi =\frac{7}{2},\quad 
    \Omega_{\eta_1} =-\frac{5}{2},\quad\Omega_{\rho_m} =0.
\end{align}

$\bullet$ \hypertarget{FBbar5}{$\bar{\mathcal{F}}_5^{(B)}$} the second solution of the system with eliminating $X_7$ by Eq.(\ref{eq:contBs}) is given by
\begin{align}
    X_2 &=0,\quad X_9=0,\nonumber\\
    X_1 &=-\sqrt{\frac{1}{373} \left(23 \sqrt{537}+81\right)} \lambda,\nonumber\\
    X_5 &=-\sqrt{\frac{1}{746} \left(47 \sqrt{537}+1171\right)} \lambda,\nonumber\\
    X_6 &=-\sqrt{\frac{1}{373} \left(23 \sqrt{537}+81\right)} \lambda.
\end{align}
The eigenvalues of the system are very complicated and we need to study the stability numerically. The result reflects that this critical point is always stable node. The effective equation of state is found to be
\begin{equation}
w_\text{eff}=-\frac{5}{3}.
\end{equation}
As a result, this critical point can be DE dominated universe at the late time
and the density parameters are found as
\begin{align}
    \Omega_A &=\frac{1}{2} \left(\sqrt{537}-33\right),\quad \Omega_\phi =\frac{1}{4} \left(\sqrt{537}+17\right),\nonumber \\
    \Omega_{\eta_1} &=\frac{1}{8} \left(43-3 \sqrt{537}\right),\quad \Omega_{\rho_m} =0.
\end{align}

\subsubsection{$\bar{\mathcal{F}}_i^{(C)}$ : $(\phi,\eta_0,\eta_1,\rho_m)$-system (C)}
We consider $(\phi,\eta_0,\eta_1,\rho_m)$-system, we assume that $X_1=0$ $,X_3=0$, $X_4=0$ and $X_6=0$. The constraint equation takes the following form,
\begin{equation}
    1 = X_2^2 + X_5^2 X_7^2 + X_5 X_9^2 - \lambda^2X_7^2- \lambda^2X_8^2.  \label{eq:contCs}
\end{equation}

$\bullet$ \hypertarget{FCbar1}{$\bar{\mathcal{F}}_1^{(C)}$}, we firstly get rid of $X_2$ from the system by the constraint equation (\ref{eq:contCs}), the critical point is given by
\begin{align}
X_5 = \frac{\sqrt{31} \lambda }{5}, \quad X_7 = \frac{5}{\sqrt{6} \lambda },\quad X_8 =0,\quad X_9=0.
\end{align}
The eigenvalues of this critical point are found as
\begin{equation}
\left(0,0,-\frac{24}{5},-\frac{3}{2}\right).
\end{equation}
The result of the eigenvalues clearly shows that the critical point is stable node. The effective equation of state is read 
\begin{equation}
    w_\text{eff}=-1.
\end{equation}
This means that this critical point can be used to describe de-Sitter DE dominated at the late time. While the density parameters are given by
\begin{align}
    \Omega_\phi =\frac{31}{6},\quad
    \Omega_{\eta_0} =0,\quad
    \Omega_{\eta_1} =-\frac{25}{6},\quad
    \Omega_{\rho_m} =0.
\end{align}

$\bullet$ \hypertarget{FCbar2}{$\bar{\mathcal{F}}_2^{(C)}$}, we exclude the $X_7$ from the autonomous system in this case. The critical point is found to be 
\begin{align}
X_2 =0,\quad X_8=0,\quad X_9=0,\quad X_5 = \sqrt{\frac{7}{5}}\lambda.
\end{align}
Noted that the critical point exists for all positive $\lambda$ parameter. The real part of the eigenvalues is given by
\begin{equation}
\left(-\frac{1}{2},1,-1-i \sqrt{\frac{37}{5}},-1+i \sqrt{\frac{37}{5}}\right)\quad \stackrel{{\rm real \ part}}{\longrightarrow}\quad \left(-\frac{1}{2},1,-1,-1\right).
\end{equation}
The result explicitly shows that this critical point is saddle point of the system. The effective equation of state is equal to
\begin{equation}
    w_\text{eff}=-\frac{1}{3}.
\end{equation}
The density parameters are found that
\begin{align}
    \Omega_\phi =\frac{7}{2},\quad
    \Omega_{\eta_0} =0,\quad
    \Omega_{\eta_1} =-\frac{5}{2},\quad
    \Omega_{\rho_m} =0.
\end{align}

$\bullet$ \hypertarget{FCbar3}{$\bar{\mathcal{F}}_3^{(C)}$}, we eliminate the $X_9$ from the system by Eq.(\ref{eq:contCs}). The critical point is determined as
\begin{align}
X_2 =0,\quad
X_5 = \sqrt{\frac{5}{3}}\lambda, \quad
X_7 = \frac{3}{\sqrt{2} \lambda }, \quad
X_8 = 0.
\end{align}
The eigenvalues of this critical point are very complicated. Then, the numerical calculation is required and the result shows that the critical point is stable for all positive $\lambda$ and $\mu$. The effective equation of state is
\begin{equation}
    w_\text{eff}=-1,
\end{equation}
which can be interpreted as late time dominated DE phase. The density parameters of this critical point are given by
\begin{align}
    \Omega_\phi =\frac{15}{2},\quad
    \Omega_{\eta_0} =0,\quad
    \Omega_{\eta_1} =-\frac{9}{2},\quad
    \Omega_{\rho_m} =-2.
\end{align}

\subsubsection{$\bar{\mathcal{F}}_i^{(D)}$ : $(\phi,\eta_0,\eta_1,A,\rho_m)$-system (D)} 
In this case, all fields are considered except the kinetic terms of $\eta_0$ and $\eta_1$, i.e. $X_4=0$ and $X_3=0$. The constraint equation reads
\begin{equation}
    1 = X_2^2 - \frac{X_1^2}{2 X_5^2} + X_5^2 X_7^2 - X_6^2 X_7^2 + X_5 X_9^2 - \lambda^2X_7^2 - \lambda^2X_8^2. \label{eq:contDs}
\end{equation}

$\bullet$ \hypertarget{FDbar1}{$\bar{\mathcal{F}}_1^{(D)}$}, we firstly replace the $X_1$ by constraint equation (\ref{eq:contDs}). The critical point of this system is given by
\begin{align}
    X_2 =&0,\quad  X_8=0,\quad X_9=0,\nonumber\\
    X_5 =& \sqrt{\frac{1}{458} \left(29 \sqrt{273}+547\right)} \lambda ,\nonumber \\
    X_6 =& \sqrt{\frac{2}{229} \left(7 \sqrt{273}-18\right)} \lambda , \nonumber\\
    X_7 =& \sqrt{\frac{3 \sqrt{273}}{8 \lambda ^2}-\frac{25}{8 \lambda ^2}},\nonumber\\
\end{align}
In addition, the critical point exists for all positive $\lambda$ and $\mu$. According to the complicated form of the eigenvalues, the numerical analysis demonstrates that this critical point is stable. The effective equation of state is equal to
\begin{equation}
    w_\text{eff} = -1,
\end{equation}
which corresponds to the DE phase dominated at the late time. The density parameters of the system are given by
\begin{align}
    \Omega_A &=\frac{1}{2} \left(\sqrt{273}-21\right), \nonumber\\
    \Omega_\phi &=\frac{1}{2} \left(\sqrt{273}+11\right), \nonumber\\
    \Omega_{\eta_1} &=\frac{1}{8} \left(25-3 \sqrt{273}\right), \nonumber\\
        \Omega_{\eta_0} &=0,\quad\Omega_{\rho_m} =0 .
\end{align}

$\bullet$ \hypertarget{FDbar2}{$\bar{\mathcal{F}}_2^{(D)}$}, getting rid of the $X_2$ from the system via constraint equation (\ref{eq:contDs}), the critical point is found as follow
\begin{align}
    X_1 &=0, \quad X_6=0,\quad X_8=0,\quad X_9=0 ,\quad 
    X_5 =\frac{\sqrt{31}\lambda}{5} ,\quad X_7=\frac{5}{\sqrt{6}\lambda}.
\end{align}
The eigenvalues of the critical point are equal to 
\begin{align}
\left(0,0,-\frac{24}{5},-\frac{3}{2},-1,1\right),
\end{align}
which come with all positive $\lambda$ and $\mu$. The effective equation of state is given by
\begin{equation}
    w_\text{eff}=-1.
\end{equation}
Although this critical point behaves as DE due to its equation of state but it does not dominate the universe at the late time. While the density parameters are read
\begin{align}
    \Omega_A =0, \quad \Omega_\phi =\frac{31}{6}, \quad
    \Omega_{\eta_0} =0, \quad
    \Omega_{\eta_1} =-\frac{25}{6}, \quad
    \Omega_{\rho_m} =0 .
\end{align}

$\bullet$ \hypertarget{FDbar3}{$\bar{\mathcal{F}}_3^{(D)}$}, we eliminate the $X_7$ from the system by using Eq.(\ref{eq:contDs}). The critical point of this system equals to
\begin{align}
    X_1=0,\quad X_2=0,\quad X_6=0,\quad X_8=0,\quad X_9=0,\quad X_5 =\sqrt{\frac{7}{5}} \lambda.
\end{align}
The eigenvalues of the critical point are
\begin{align}
\left(0,-\frac{9}{5},-1,-\frac{1}{2},1,2\right).
\end{align}
This reflects that this critical point is always saddle point. The effective equation of state is given by
\begin{equation}
    w_\text{eff}=-\frac{1}{3}.
\end{equation}
The density parameters of this system are written by
\begin{align}
    \Omega_A =0, \quad \Omega_\phi=\frac{7}{2}, \quad \Omega_{\eta_0} =0,\quad \Omega_{\eta_1} =-\frac{5}{2}, \quad \Omega_{\rho_m} =0 .
\end{align}

$\bullet$ \hypertarget{FDbar4}{$\bar{\mathcal{F}}_4^{(D)}$}, as done previously for eliminating $X_7$, the critical point reads
\begin{align}
X_2 &=0, \quad X_8=0,\quad X_9=0,\nonumber \\
    X_1 &=\sqrt{\frac{1}{373} \left(23 \sqrt{537}+81\right)} ,\nonumber\\
    X_5 &=\sqrt{\frac{1}{746} \left(47 \sqrt{537}+1171\right)},\nonumber\\
    X_6 &=\sqrt{\frac{1}{373} \left(23 \sqrt{537}+81\right)} \lambda.
\end{align}
The stability analysis of this critical point is performed numerically and the result shows that this critical point is stable for all positive $\lambda$. While the effective equation of state is equal to
\begin{equation}
    w_\text{eff}=-\frac{5}{3}.
\end{equation}
The density parameters of this system are given by
\begin{align}
    \Omega_A &=\frac{1}{2} \left(\sqrt{537}-33\right), \nonumber\\
    \Omega_\phi &=\frac{1}{4} \left(\sqrt{537}+17\right), \nonumber\\
    \Omega_{\eta_1} &=\frac{1}{8} \left(43-3 \sqrt{537}\right), \nonumber\\
        \Omega_{\eta_0} &=0, \quad \Omega_{\rho_m} =0 .
\end{align}
\section{Discussion}\label{sec:Discussion}
From the previous section, we obtained a number of critical points from KK inspied BD gravity model with barotropic matter. The critical points are classified by their stability. Then the effective equation of state for each critical point are calculated. In analysis, the stability and effective equation of state are used to interpret the physical meaning behind each critical point. Consequently, we acquired critical points which are both the compatible and incompatible with the observable universe.



For the DM, there are 2 critical points which responsible for it. The first one comes from $\mathcal{F}_i^{(B)}$ system where non-vanishing fields are $\phi, \eta_1, A^\mu, \rho_m$, i.e., \hyperlink{FB1}{$\mathcal{F}_1^{(B)}$}, while the other one comes from the $\mathcal{F}_i^{(D)}$ system, i.e., \hyperlink{FD1}{$\mathcal{F}_1^{(D)}$}, where all fields are not vanished. Apparently, these two critical points are the same critical point coming from different system. Since the effective equation of state of these critical points is zero, i.e. $w_\text{eff}=0$, and their stability are saddle, these critical points can be represented as matter-dominated phase in the standard cosmological model. Noted that the matter-dominated phase is contributed by both matter and DM, then DM behaviour is obtainable from our model.

For the DE, we have 4 critical points from 3 systems, namely, $\bar{\mathcal{F}}_i^{(B)}$, $\bar{\mathcal{F}}_i^{(C)}$ and $\bar{\mathcal{F}}_i^{(D)}$. The first point is \hyperlink{FBbar3}{$\bar{\mathcal{F}}_1^{(B)}$} where $\phi, \eta_1, A^\mu, \rho_m$ are non-vanishing fields. The second and third points are \hyperlink{FCbar1}{$\bar{\mathcal{F}}_1^{(C)}$} and \hyperlink{FCbar3}{$\bar{\mathcal{F}}_3^{(C)}$} where  $\phi, \eta_0, \eta_1, A^\mu, \rho_m$ are non-vanishing fields. The fourth point is \hyperlink{FDbar1}{$\bar{\mathcal{F}}_1^{(D)}$} where all fields are not vanished. Noted that for barred system we assumed that the fields $\eta_0$ and $\eta_1$ are slow-roll fields causing the dynamics of these fields to be zero, i.e. $\dot{\eta}_0=\dot{\eta}_1=0$. These 4 critical points are all stable points and their effective equations of state are simply $-1$, $w_\text{eff}=-1$ which give us accelerating universe driven by DE. This means that our model is sufficient to predict DE phase.

Moreover, we also obtain the phantom DE solutions, i.e. $w_\text{eff}<-1$. This solution gives us the highly accelerated expanding universe leading to the Big Rip. We have 3 suitable critical points for the phantom DE. The first one is \hyperlink{FB2}{$\mathcal{F}_2^{(B)}$} where $\phi, \eta_1, A, \rho_m$ are non-vanishing fields. This critical point will be stable point with effective equation of state, $w_\text{eff}<-5/3$ for $\lambda > 3/\mu$. The second one is \hyperlink{FDbar5}{$\bar{\mathcal{F}}_5^{(D)}$} where non-vanishing fields are the same as \hyperlink{FB2}{$\mathcal{F}_2^{(B)}$} excepting that $\eta_1$ is a slow-roll field in this system, i.e. $\dot{\eta}_1=0$. The critical point is stable point with effective equation of state, $w_\text{eff}=-5/3$. The third one is \hyperlink{FDbar4}{$\bar{\mathcal{F}}_4^{(D)}$} where all fields are not vanished. For this critical point, we assumed that $\dot{\eta}_0$ and $\dot{\eta}_1$ are slow-roll fields. This critical point appears to be a stable point and its effective equation of state is less than $-5/3$, i.e., $w_\text{eff}=-5/3$.


For the latter, we also obtain critical points which incompatible with the observable universe. Now, we are getting into details of those critical points. Starting from \hyperlink{FBbar4}{$\bar{\mathcal{F}}_4^{(B)}$}, \hyperlink{FCbar2}{$\bar{\mathcal{F}}_2^{(C)}$} and \hyperlink{FDbar3}{$\bar{\mathcal{F}}_3^{(D)}$}, they are all saddle point. The effective equations of state for these critical points are perfectly stand on the border of decelerated and accelerated expansion of the universe, i.e., $w_\text{eff}=-1/3$. This suggests constantly expanded universe and we will call this situation as critical DE phase. Now, for \hyperlink{FA1}{$\mathcal{F}_1^{(A)}$}, \hyperlink{FA2}{$\mathcal{F}_2^{(A)}$} and \hyperlink{FC1}{$\mathcal{F}_1^{(C)}$}, they are stable points and their effective equation of state, $w_\text{eff}$, is zero. This kind of critical point represents the universe evolving to matter-dominated phase and remaining at this phase forever. 
Next, \hyperlink{FBbar2}{$\bar{\mathcal{F}}_2^{(B)}$}, \hyperlink{FBbar3}{$\bar{\mathcal{F}}_3^{(B)}$} and \hyperlink{FDbar2}{$\bar{\mathcal{F}}_2^{(D)}$}, they are saddle points with DE’s equation of state, i.e. $w_\text{eff}=-1$. The universe, according to the properties of these critical points, is evolving to DE dominated era. Then, it will transit to unpredictable phase. Lastly \hyperlink{FB3}{$\mathcal{F}_3^{(B)}$} and \hyperlink{FD2}{$\mathcal{F}_2^{(D)}$}, they are also saddle points with effective equation of state related to phantom DE, i.e. $w_\text{eff}=-5/3$. This predicts that the phantom DE is not the final phase of the universe. The universe will continually evolve to the unknown phase after.

\section{\label{sec:Conclusion and Outlook}Conclusion and Outlook}

In this paper, we have revisited the KK inspired BD model where barotropic fluid is included. The UV limit of this model is the traditional 5-dimensional KK action with 1 additional massive scalar field and 2 gauge fields. The KK compactification reduces the UV theory to a 4-dimensional gravity theory with a dilaton coupling to a tower of scalar fields. Another relic of the higher dimensions is the gauge field from 5-dimensional metric which forms a Cosmic Triad solution with 2 additional gauge fields. Then the usual Einstein field equations with the barotropic fluid are calculated. Together with equations of motion, the complete autonomous system is constructed and the appropriate dimensionless parameters are defined.

With the dynamical system approach, we found that realistic DM and DE phases are readily present in many solutions. Since our universe has undergone the matter dominated phase and currently it is in the DE dominated phase, the DM solutions are expected to be saddle points with $w_{\rm eff} = 0$. In the analysis we found 1 DM solution, i.e,  \hyperlink{FB1}{$\mathcal{F}_1^{(B)}$} (and equivalently \hyperlink{FD1}{$\mathcal{F}_1^{(D)}$}). The \hyperlink{FB1}{$\mathcal{F}_1^{(B)}$} critical point consists of the oscillating $\dot{\eta}_1$ field and ($\phi$, $A^{\mu}$, $\rho_m$) which play a role of background fields. We found 4 dark energy solutions which are stable critical points with $w_{\rm eff} = -1$, i.e.,  \hyperlink{FBbar3}{$\bar{\mathcal{F}}_1^{(B)}$}, \hyperlink{FCbar1}{$\bar{\mathcal{F}}_1^{(C)}$}, \hyperlink{FCbar3}{$\bar{\mathcal{F}}_3^{(C)}$} and \hyperlink{FDbar1}{$\bar{\mathcal{F}}_1^{(D)}$}. Noted that all of these are in the slow-roll approximation as one might expected. Interestingly, we also found many solutions which resemble the phantom DE with $w_{\rm eff} < -1$. These are \hyperlink{FB2}{$\mathcal{F}_2^{(B)}$}, \hyperlink{FDbar5}{$\bar{\mathcal{F}}_5^{(D)}$}, and \hyperlink{FDbar4}{$\bar{\mathcal{F}}_4^{(D)}$}. All of these solutions contain non-zero kinetic terms which suggests that the phantom DE could change the value along the history of the universe.

Although this approach could not provide the origin of the barotropic fluid at the 5-dimensional level, we consider this approach as an effective analysis of the overall contributions of other fields. The lack of a strong connection to UV physics is compensated by the fact that the revisited model presented here has many more facets of the DM/DE behaviours comparing to the original model. One could conclude that the barotropic fluid in the KK inspired BD model plays an important role in enriching phenomena of the model tremendously.


\begin{acknowledgments}
AW acknowledges the support of the Development and Promotion of Science and Technology Talents Project (DPST), the Institute for the Promotion of Teaching Science and Technology (IPST). The work of TK, CP and DS has been supported by the National Astronomical Research Institute of Thailand. DS is supported the Mid-Career Research Grant 2021 from National Research Council of Thailand.
\end{acknowledgments}

\bibliography{KK-BD}


\end{document}